\documentclass[12pt]{iopart}

\usepackage{iopams}
\usepackage{cite}
\usepackage{graphicx}
\usepackage{amssymb}
\usepackage{textcomp}
\usepackage{multirow}
\usepackage{booktabs}
\usepackage[british]{babel}
\usepackage{color}
\usepackage[T1]{fontenc}

\begin{document}

\title{Universal spectral features of different classes of random diffusivity
processes}

\author{Vittoria Sposini$^{\dagger,\ddagger}$, Denis S Grebenkov$^{\flat}$, Ralf
Metzler$^{\dagger}$, Gleb Oshanin$^{\mathsection, \mathparagraph}$ and Flavio Seno$^{\|}$}
\address{$\dagger$ Institute for Physics \& Astronomy, University of Potsdam,
14476 Potsdam-Golm, Germany}
\address{$\ddagger$ Basque Centre for Applied Mathematics, 48009 Bilbao, Spain}
\address{$\flat$ Laboratoire de Physique de la Mati\`{e}re Condens\'ee (UMR 7643),
CNRS -- Ecole Polytechnique, IP Paris, 91128 Palaiseau, France}
\address{$\mathsection$ Sorbonne Universit\'{e}, CNRS, Laboratoire de Physique Th\'{e}orique
de la Mati\`{e}re Condens\'{e}e (UMR 7600), 4 Place Jussieu, 75252 Paris Cedex 05,
France}
\address{$\mathparagraph$ Interdisciplinary Scientific Center J.-V. Poncelet (ISCP), CNRS UMI 2615, 11 Bol. Vlassievsky per., 119002 Moscow, Russia}
\address{$\|$ INFN, Padova Section and Department of Physics and Astronomy
"Galileo Galilei", University of Padova, 35131 Padova, Italy}
\ead{rmetzler@uni-potsdam.de (corresponding author)}

\begin{abstract}
Stochastic models based on random diffusivities, such as the diffusing-diffusivity
approach, are popular concepts for the description of non-Gaussian diffusion in
heterogeneous media. Studies of these models typically focus on the moments and
the displacement probability density function. Here we develop the complementary
power spectral description for a broad class of random diffusivity processes. In
our approach we cater for typical single particle tracking data in which a small
number of trajectories with finite duration are garnered. Apart from the
diffusing-diffusivity model we study a range of previously unconsidered random
diffusivity processes, for which we obtain exact forms of the probability density
function. These new processes are different versions of jump processes as well as
functionals of Brownian motion. The resulting behaviour subtly depends on the
specific model details. Thus, the
central part of the probability density function may be Gaussian or non-Gaussian,
and the tails may assume Gaussian, exponential, log-normal or even power-law forms.
For all these models we derive analytically the moment-generating function for the
single-trajectory power spectral density. We establish the generic $1/f^2$-scaling
of the power spectral density as function of frequency in all cases. Moreover, we
establish the probability density for the amplitudes of the random power
spectral density of individual trajectories. The latter functions reflect the very
specific properties of the different random diffusivity models considered here.
Our exact results are in excellent agreement with extensive numerical simulations.
\end{abstract}

\today

\section{Introduction}

Diffusive processes came to the attention of the broader scientific community
with the experiments on "active molecules" by Robert Brown, who reported the
jittery motion of granules of "1/4000th to 1/5000th of an inch in length"
contained in pollen grains as well as control experiments on powdered inorganic
rocks \cite{brown}. In the mid-19th century physician-physiologist Adolf Fick
published his studies on salt fluxes between reservoirs of different
concentrations connected by tubes \cite{fick}. To quantify the observed
dynamics Fick introduced the diffusion equation ("Fick's second law") for the
spatio-temporal concentration profile. A major breakthrough was the theoretical
description of "Brownian motion" and the diffusion equation in terms of
probabilistic arguments by Einstein \cite{einstein}, Smoluchowski
\cite{smoluchowski}, and Sutherland \cite{sutherland}. Concurrently Karl
Pearson introduced the notion of the "random walk" \cite{pearson}, and
Langevin proposed the intuitive picture of the random force and the stochastic
Langevin equation \cite{langevin}.

More recently, major advances in experimental techniques such as superresolution
microscopy continue to provide unprecedented insight into the motion of submicron
and even molecular tracers in complex environments such as living biological cells
\cite{hoefling,lene,xie,brauchle}. Concurrently, simulations are becoming ever
more powerful and reveal the Molecular Dynamics in systems such as lipid membranes
\cite{ilpo} or internal protein motion \cite{smith}. The data resulting
from such complex systems unveil a number of new phenomena in the stochastic
particle motion and thus call for new theoretical concepts \cite{pt,pt1,
statphys} on top of already known approaches \cite{yossipt,mikenature,harveypt}.

Among these new insights is that endogenous and introduced tracers in living
biological cells perform anomalous diffusion of the form $\langle\textbf{r}^2(t)
\rangle\simeq K_{\alpha}t^{\alpha}$ in a wide range of systems \cite{hoefling,
lene,saxton}. For instance, subdiffusion with $0<\alpha<1$ was measured for
messenger RNA probes in bacteria cells \cite{golding,weber}, for DNA loci and
telomeres in bacteria and eukaryotic cells \cite{weber,yuval,yuval1}, for granules
in yeast and human cells \cite{leneprl,greb}, as well as for the stochastic motion of
biological membrane constituents \cite{diego,carlo}. In these cases the slower than
Brownian, passive tracer motion is effected by the highly crowded nature of the
environment, as can be studied in \emph{in vitro\/} systems \cite{matthias,lene1}.
In fact, even small green fluorescent proteins of some 2 nanometre in size
were shown to subdiffuse \cite{gratton}. Conversely, superdiffusion with $1<
\alpha<2$ in biological cells is caused by active motion of molecular motors
due to consumption of biochemical energy units. Examples include the motor
motion itself \cite{granick,roberts}, the transport of introduced plastic
beads in fibroblast cells \cite{elbaum}, RNA cargo in neuron cells
\cite{jaenature}, or of granules in amoeba \cite{christine}.

However, even when the mean squared displacement seemingly suggests Brownian
motion based on the observation that $\alpha=1$, remarkable effects have been
reported recently. Thus,
the motion of micron-sized tracer beads moving along nanotubes as well as in
entangled polymer networks was shown to be "Fickian" yet the
measured displacement distribution exhibited significant deviations from the
expected Gaussian law: namely, an exponential distribution of the form
$P(\mathbf{r},t)\propto\exp(-|\mathbf{r}|/\lambda(t))$ with $\lambda(t)\propto
t^{1/2}$ was observed \cite{granicknm,granick1}. Similar "Fickian yet non-Gaussian"
diffusion was found for the tracer dynamics in hard sphere colloidal
suspensions \cite{granick2}, for the stochastic motion of nanoparticles in
nanopost arrays \cite{post}, of colloidal nanoparticles adsorbed at fluid
interfaces \cite{xue:BYNG8,xue1,xue2} and moving along membranes and inside colloidal
suspension \cite{Goldstein:BYNG9}, and for the motion of nematodes \cite{hapca}.
Even more complicated non-Gaussian distributions of displacements were recently
observed in \emph{Dictyostelium discoideum\/} cells \cite{greb1,beta} and
protein-crowded lipid bilayer membranes \cite{ilpo1}.
While in some experiments the non-Gaussian shape of $P(\mathbf{r},t)$ is
observed over the entire experimental window, others report clear crossover
behaviours from a non-Gaussian shape at shorter time scales to an effective
Gaussian behaviour at longer time scales, for instance, see \cite{granicknm,
granick1}. 

A non-Gaussian probability density along with the scaling exponent $\alpha=1$
of the mean squared displacement can be achieved in the superstatistical
approach, in which it is assumed that individual Gaussian densities are
averaged over a distribution of diffusivities \cite{beck,beck1,beck2,beck3,erice}.
A microscopic realisation of such a behaviour was proposed for a model of diffusion
during a polymerisation process \cite{fulvio}.
However, in superstatistics (and in the related process called generalised
grey Brownian motion \cite{gianni,gianni1,gianni2}) the distribution is a constant
of the motion and thus no crossover behaviour as mentioned above can be described.
In order to include such a non-Gaussian to Gaussian crossover models were
introduced in which the diffusion coefficient is considered as a stochastic
process itself. In this diffusing-diffusivity picture, originally proposed
by Chubynsky and Slater \cite{Chubynsky14}, the stochastic dynamics of the
diffusivity is characterised by a well-defined correlation time above which
the diffusivity becomes equilibrated. Concurrently to this equilibration
the ensuing form of $P(\mathbf{r},t)$ becomes effectively Gaussian. Random
diffusivity models have since then been developed and analysed further, and
their application is mainly the diffusive dynamics in heterogeneous systems
\cite{Jain16,Jain16b,Chechkin:DD2,cherayil,Lanoiselee18a,Lanoiselee18b,Sposini18,Grebenkov19,Lanoiselee19,Sposini19,Barkai19,Roichman19}.

In fact, stochastic models based on random diffusivities are ubiquitous in
financial mathematics for the modelling of stock price.  They are commonly
known as stochastic volatility models and many different examples have
been analysed in order to identify a proper description for the volatility
\cite{Shephard}. Among them, one can find diffusion-based models, where
the volatility is described with continuous sample paths, as well as more
complicated dynamics where, for instance, jumps are also allowed or where
the volatility is defined as a function of separate stochastic processes
\cite{Barndorff-Nielsen}. Financial mathematics hosts a rich variaty of
random diffusivity models, motivated by various aspects of the observed
financial market data. Here we present a range of additional, new random
diffusivity models in the context of the time series analysis, generalising
the diffusing-diffusivity model developed for Fickian yet non-Gaussian
diffusion processes. As this field is quickly expanding and new facets are
being continuously unveiled, we are confident that these different models
and their features offer the necessary flexibility to account for these new
observations.

The central purpose of our study here is twofold. First we analyse several new
classes of random diffusivity models, divided into two groups, jump models and
functionals of Brownian motion. For both groups we consider several concrete
examples and derive analytic solutions for the probability density function (PDF)
${\bf{\bf \Pi}}(x,t)$. The PDF turns out to delicately depend on the precise formulation
of the model: the central part may be Gaussian or non-Gaussian, and the tails
may be of Gaussian, exponential, log-normal, or even power-law shape. The second
goal we pursue here are the \emph{spectral\/} properties of random diffusivity
processes. Namely, while earlier studies of the random diffusivity dynamics
were mainly concerned with the PDF and the mean squared
displacement encoded in the process we assume a different stance and derive the
the spectral properties of single particle trajectories with finite observation
time, geared for the description of contemporary single particle tracking
experiments. Such an analysis was worked out in detail for specific systems of
normal and anomalous diffusion \cite{Krapf14,Barkai16,Barkai17,we1,we2,vittoria},
and we here study the commonalities and differences emerging for random diffusivity
scenarios.

Traditionally, power spectral analyses are based on the textbook definition
of the spectral density
\begin{equation}
\mu(f)=\lim_{T\to\infty}\frac{1}{T}\left\langle\left|\int_0^Te^{ift}x(t)dt
\right|^2\right\rangle.
\end{equation}
This definition involves taking the limit of infinite (practically, very long)
measurement times as well as averaging over an ensemble (practically, a large
number of) of particles, here and in the following denoted by angular brackets,
$\langle\cdot\rangle$. Typical single particle tracking experiments, however,
are limited in the measurement time, for instance, due to the lifetime of
the employed fluorescent tags or the time a particle stays in the microscope
focus.  At the same time, such experiments are often limited to a relatively
small number of individual trajectories. To cater for this common type of
experimental situations we avoid taking the long time and ensemble limits
by considering the single-trajectory power spectral density (PSD)
\begin{equation}
S_T(f)=\frac{1}{T}\left|\int_0^Te^{ift} x(t)dt\right|^2
\label{PSD}
\end{equation}
as functions of frequency $f$ and measurement time $T$. We previously analysed
the behaviour of $S(f,T)$ for different diffusion scenarios \cite{we1,we2,
vittoria} and demonstrated that it is practically useful in the analysis of
experimental data \cite{we1,we2}. In what follows we derive the moment-generating
function (MGF) of the PSD (\ref{PSD}) for different classes of random-diffusivity
processes, including several cases not yet studied in literature.
In particular, we obtain the probability density $P(A)$ of the single
PSD amplitude, an intrinsically random quantity for a finite-time
measurement of a stochastic motion that was demonstrated to be a very useful
quantity for the analysis of measured particle trajectories. In addition to
analytical derivations we present detailed numerical analyses. This study
provides a quite general approach to obtain the PDF for any diffusing
diffusivity model, providing new insights on this class of processes.

This work is structured as follows. We start from section \ref{sec1} with
a description of the model and in section \ref{sec2} we report general
results on the spectral properties of this class of processes. Specific
examples of diffusing-diffusivity models are described in sections \ref{sec3}-\ref{m}.
The first example is the well known case in which the diffusivity is modelled
as the squared Ornstein-Uhlenbeck process. In the second group of examples we
analyse two cases in which the diffusivity is defined as a jump process. The
third and last group shows three examples in which the diffusivity is described
as a functional of Brownian motion. Finally, in section \ref{sec4} we draw our
conclusions. In the appendix we report details on the explicit derivations of
our results.

\section{Random diffusivity processes}
\label{sec1}

We consider a class of one-dimensional stochastic processes $x_t$ that obey the
Langevin equation in the It\^o convention,
\begin{equation}
\label{lan}
\dot{x}_t=\sqrt{2D_0\Psi_t}\xi_t.
\end{equation}
Here $D_0$ is a constant, dimensional coefficient in units ${\rm length}^2/{\rm
time}$, and in our analysis we will assume the initial condition $x_0=0$. In
equation (\ref{lan}) $\xi_t$ denotes a standard Gaussian white noise with zero
mean and covariance function $\overline{\xi_t\xi_{t'}}=\delta\left(t-t'\right)$.
The bar here and henceforth denotes averaging with respect to the noise $\xi_t$.
Lastly $\Psi_t$ is a positive-definite random function, which multiplies $D_0$
and thus introduces a time-dependent randomness into the effective noise
amplitude. In the following we stipulate that $\Psi_t$ is Riemann-integrable on
a finite interval $(0,T)$ such that $\int^T_0dt\Psi_t$ exists with probability
$1$. Note that the case $\Psi_t\equiv1$ corresponds to standard Brownian motion,
while a deterministic choice of the form $\Psi_t=t^{\alpha-1}$ produces so-called
scaled Brownian motion \cite{sbm,sbm1}. We will here discuss several particular
choices for the random function $\Psi_t$. In addition to the previously made
choice of a squared Ornstein-Uhlenbeck process we will consider the case when
$\Psi_t$ is a jump-process, that attains independent, identically distributed
random values. We also present several examples when $\Psi_t$ is subordinated
to standard unbiased Brownian motion $B_t$: namely, $\Psi_t=B^2_t/a^2$, where
$a$ is a model parameter, $\Psi_t=\Theta(B_t)$, where $\Theta(x)$ is the
Heaviside theta function, and geometric Brownian motion $\Psi_t=\exp(-B_t/a)$.

Regardless of the choice of the random function $\Psi_t$, we can solve the
Langevin equation (\ref{lan}) for the trajectory $x_t$ for a fixed realisation
of the noise and a given realisation of $\Psi_t$, to obtain
\begin{equation}
x_t=(2D_0)^{1/2}\int^{t}_0d\tau\Psi^{1/2}_{\tau}\xi_{\tau}.
 \end{equation}
The characteristic function of $x_t$ can be written down in the form
\begin{equation}
\Phi_w=\left<\overline{\exp\left(iw(2D_0)^{1/2}\int^t_0d\tau\Psi^{1/2}_{\tau}\xi
_{\tau}\right)}\right>_{\Psi},
\end{equation}
where the bar stands for averaging over thermal histories, while the angular
brackets denote averaging over the realisations of the random function $\Psi_t$.
The thermal average can be performed straightforwardly to give
\begin{equation}
\Phi_w=\left<\exp\left(-D_0w^2\int^t_0d\tau\Psi_{\tau}\right)\right>_{\Psi}.
\label{mgf_dd}
\end{equation}
The desired PDF ${\bf \Pi}(x,t)$ can then be written as
\begin{equation}
{\bf \Pi}(x,t)=\frac{1}{2\pi}\int^{\infty}_{-\infty}dwe^{-iwx}\Phi_w.
\label{pdf_DD}
\end{equation}
In the following section \ref{sec3} we provide several examples with explicit
expressions for the probability density, and we will see how different choices
of $\Psi_t$ may lead to PDFs of considerably different shapes.

\section{General theory}
\label{sec2} 

We first obtain exact expressions for the PSD (\ref{PSD}) and then study the
limiting behaviour for high frequencies.

\subsection{Exact expressions for arbitrary frequency and observation time}

We investigate the PSD of an individual trajectory $x_t$ encoded in the
stochastic dynamics (\ref{lan}) with $t\in(0,T)$,
\begin{equation}
\label{1}
S_T(f)=\frac{1}{T}\int^T_0dt_1\int^T_0dt_2\cos\left(f(t_1-t_2)\right)x_{t_1}
x_{t_2},
\end{equation}
as function of the frequency $f$ and the observation time $T$. We determine
the MGF and the PDF of the random variable $S_T(f)$.

The MGF of the single-trajectory PSD in (\ref{1}) is
defined as
\begin{equation}
\phi_{\lambda}=\left<\overline{\exp\left(-\frac{\lambda}{T}\int^T_0dt\int^T_0
dt'\cos\left(f(t-t')\right)x_{t}x_{t'}\right)}\right>_{\Psi}
\label{b}
\end{equation}
with $\lambda\geq0$. Relegating some intermediate calculations to
\ref{app} we find the following expression for $\phi_{\lambda}$ in (\ref{b})
averaged over thermal noises, 
\begin{eqnarray}
\nonumber
\phi_{\lambda}&=&\frac{1}{4\pi\lambda}\int^{\infty}_{-\infty}dz_1\int^{\infty}_{
-\infty}dz_2\exp\left(-\frac{z_1^2+z_2^2}{4\lambda}\right)\\
&&\times\left\langle\exp\left(-D_0\int^T_0 dt\Psi_t\left(\int^T_td\tau Q_{\tau}
\right)^2\right)\right\rangle_{\Psi},
\label{4}
\end{eqnarray}
where
\begin{eqnarray}
\label{Q}
Q_t = z_1 \frac{\cos(f t)}{\sqrt{T}} + z_2 \frac{\sin(f t)}{\sqrt{T}}.
\end{eqnarray}
Performing the inverse Laplace transform of expression (\ref{4}), we find the
general result for the PDF
\begin{eqnarray}
\nonumber
p(S_T(f)=S)&=&\displaystyle\frac{1}{4\pi}\int^{\infty}_{-\infty}dz_1\int^{\infty}_{
-\infty}dz_2 J_0\left(\sqrt{\left(z_1^2 + z_2^2\right) S}\right)\\
&& \times\left\langle\exp\left(-D_0\int^T_0dt\Psi_t\left(\int^T_t
d\tau Q_{\tau}\right)^2\right)\right\rangle_{\Psi},
\label{8}
\end{eqnarray}
where $J_0(z)$ denotes the Bessel function of the first kind. A more explicit
dependence on the frequency $f$ can be obtained in the form (see
\ref{app} for more details)
\begin{eqnarray}
\nonumber
\phi_{\lambda}&=&\Bigg \langle \Bigg[ 1 + \frac{8 \lambda D_0}{f^2 T} \int^T_0 dt
\Psi_t \bigg(1-\cos\left(f\left(T-t\right)\right)\bigg)\\
&& + \frac{16 \lambda^2 D_0^2}{f^4 T^2 } \int^T_0 dt_1\Psi_{t_1} \int^T_0 dt_2
\Psi_{t_2}\left(\frac{3}{4}+L_f(t_1,t_2)\right)\Bigg]^{-1/2}
\Bigg\rangle_{\Psi},
\label{main}
\end{eqnarray}
where $L_f(t_1,t_2)$ is defined by the somewhat lengthy expression (\ref{L}).

The expression within the angular brackets in relation (\ref{main}) is the exact
MGF of the PSD of the process $x_t$ in (\ref{lan}) for any fixed realisation of
$\Psi_t$ and holds for arbitrary $T$ and arbitrary $f$. It also represents the
exact form of the MGF in the case when $\Psi_t$ is non-fluctuating: in particular,
for $\Psi_t=1$ it describes the MGF in case of standard Brownian motion \cite{we1},
while the choice $\Psi_t=t^{\alpha-1}$ corresponds to the case of scaled Brownian
motion recently studied in \cite{vittoria}.

\subsection{Exact high frequency limiting behaviour}

As already remarked we here concentrate on random processes $\Psi_t$ which, for
any finite $T$, are Riemann-integrable with probability $1$, which implies that
in the limit $f\to\infty$ certain integrals vanish, as shown in \ref{B}. 
As a consequence, expression (\ref{main}) attains the following exact analytic
high-frequency form
\begin{equation}
\phi_{\lambda}\sim\Bigg\langle\Bigg[1+\frac{8\lambda D_0}{f^2T}\int^T_0dt\Psi_t
+\frac{12 \lambda^2 D_0^2}{f^4 T^2 } \left(\int^T_0 dt \, \Psi_{t}\right)^2 
\Bigg]^{-1/2} \Bigg \rangle_{\Psi},
\label{mainfinf}
\end{equation}
in which we dropped the vanishing terms and kept only the leading terms in $1/f$.

We note that the Laplace parameter $\lambda$ appears in the combination $D_0
\lambda/f^2$ so that the high-$f$ spectrum of a single-trajectory PSD has the
universal form
\begin{equation}  
\label{eq:ST_A}
S_T(f)\sim\frac{4D_0A}{f^2},
\end{equation}
regardless of the specific choice of $\Psi_t$. Here $A$ is a dimensionless, random
amplitude, which differs from realisation to realisation. This means
that the characteristic high-frequency dependence of the PSD can be learned, in
principle, from just a single trajectory, in agreement with the conclusions in
\cite{we1,we2, vittoria}.

The MGF $\Phi_{\lambda}$ of the random amplitude $A$
follows from (\ref{mainfinf}) and can be written as
\begin{eqnarray}
\nonumber
\Phi_{\lambda}&=&\int\limits_0^\infty dAe^{-\lambda A}P(A)\\
&=&\frac{2}{\sqrt{3}}\int^{\infty}_0dp\exp\left(-\frac{4p}{3}\right)
I_0\left(\frac{2p}{3}\right)\Upsilon(T;\lambda p/T),
\label{mainfinf3}
\end{eqnarray}
where $I_0(z)$ is the modified Bessel function of the first kind, and
\begin{equation}
\Upsilon(T; \lambda)=\Bigg\langle\exp\left(-\lambda\int^T_0dt\Psi_t\right)
\Bigg\rangle_{\Psi} 
\label{mgf_id}
\end{equation}
is the MGF of the integrated diffusivity (see
\cite{Lanoiselee18b, Grebenkov19})
\begin{equation} 
\label{tau}
\tau_T=\int\nolimits_0^Tdt\Psi_t.  
\end{equation}
Relation (\ref{mainfinf3}) links the MGFs of $A$ and $\tau_T$. Moreover,
note that the characteristic function of the
diffusing-diffusivity model in (\ref{mgf_dd}) is tightly related to the
MGF of $\tau_T$ in (\ref{mgf_id}), specifically
\begin{equation}
\Upsilon(T; D_0 w^2)=\Phi_w.
\end{equation}
As shown in \cite{Lanoiselee18b} the function $\Upsilon(T;\lambda)$ determines
the first-passage time properties of the stochastic process $x_t$. Here we show
how this function controls the high-frequency behaviour of the PSD.

Taking the inverse Laplace transform with respect to the parameter $\lambda$
we evaluate the PDF of $A$,
\begin{eqnarray}
\nonumber
P(A)&=&\frac{2}{\sqrt{3}}\int^{\infty}_0dzJ_0\left(\left(1+\frac{1}{\sqrt{3}}\right)  
\sqrt{2zA}\right)\\
&&\times J_0\left(\left(1 - \frac{1}{\sqrt{3}}\right)  \sqrt{2 z A} \right) 
\Upsilon(T; z/T).
\label{distlargef3}
\end{eqnarray}
This exact expression determines the high-$f$ behaviour of the PDF $p(S_T(f)=S)$,
\begin{equation}
p(S_T(f)=S)\sim\frac{f^2}{4D_0}P\left(A=\frac{Sf^2}{4D_0}\right)
\end{equation}
as $f\to\infty$. The exact high-$f$ forms in (\ref{mainfinf3}) and
(\ref{distlargef3}) will serve as the basis of our analysis for several
particular choices of the process $\Psi_t$ in section \ref{sec3}.

Before proceeding, we stop to make several general statements.

(i) Expanding the exponential function on the right hand side of
(\ref{mainfinf3}) into the Taylor series in powers of $\lambda$ we obtain
straightforwardly the relation between the moments of $A$ and the moments
of the integrated diffusivity $\tau_T$, which is valid for any $n$,
\begin{equation}
{\mathbb E}\{ A^n\}=\left(\frac{3}{4}\right)^{n+1/2}n! {_2F_1}\left(\frac{n+1}{2},\frac{n+2}{2};1;\frac{1}{4}\right)\left \langle\frac{\tau_T^n}{T^n}\right\rangle_{\Psi},
\label{moments3}
\end{equation}
where $_2F_1(a,b;c;z)$ is the Gauss hypergeometric function. 
Since the moments of $\tau_T$ are related to the moments of the process $x_T$
\cite{Grebenkov19} we also find
\begin{equation}
{\mathbb E}\{ A^n\}=\left(\frac{3}{4}\right)^{n+1/2}\frac{(n!)^2}{(2n)!}
\,_2F_1\left(\frac{n+1}{2}, \frac{n+2}{2}; 1; \frac{1}{4}\right)\\
\left \langle \frac{x_T^{2n}}{(D_0T)^n} \right \rangle_{\Psi}.
\label{moments3a}
\end{equation}

(ii) Starting from the results in (\ref{moments3}) and (\ref{moments3a}) we can 
readily obtain the moments of $S_T(f)$ as well. In particular, if we focus on
its average value, we have 
\begin{equation}
\langle S_T(f)\rangle=\frac{4D_0C_1}{f^2T}\langle\tau_T\rangle_\Psi=\frac{2C_1}{
f^2T}\left\langle\overline{x_T^2}\right\rangle_\Psi,
\label{muT}
\end{equation}
where $C_1={(3/4)^{3/2}}_2F_1(1,3/2;1;1/4)$. This suggests that those random
diffusivity models that display anomalous scaling of the MSD, i.e., $\left
\langle\overline{x_T^2}\right\rangle_\Psi\not\simeq T$, exhibit ageing behaviour,
namely, a dependence of the PSD properties on the trajectory length $T$.

(iii) Equations (\ref{eq:ST_A}) and (\ref{moments3}) permit us to directly
access the coefficient of variation $\gamma$ of the PDF $p(S_T(f)=S)$ in the
high-$f$ limit. We get straightforwardly
\begin{eqnarray}
\nonumber
\gamma&=&\left(\frac{\left\langle\overline{S_T^2(f)}\right\rangle_{\Psi}-
\left \langle \overline{S_T(f)} \right \rangle^2_{\Psi} }{\left \langle
\overline{S_T(f)} \right \rangle^2_{\Psi} } \right)^{1/2}
\approx\left(\frac{\left\langle A^2\right\rangle_{\Psi}-\left\langle A\right
\rangle^2_{\Psi}}{\left \langle A \right \rangle^2_{\Psi} } \right)^{1/2}\\
\nonumber
&=&\left(\frac{3}{4}\frac{\left\langle x^4_T \right \rangle_{\Psi}}{\left\langle
x^2_T \right \rangle^2_{\Psi}}  - 1\right)^{1/2}
=\left(\frac{9}{4} \frac{\left \langle \tau^2_T \right \rangle_{\Psi}}{\left
\langle \tau_T \right \rangle^2_{\Psi}}  - 1\right)^{1/2}\\
&=&\left(\frac{9}{4} 
\frac{\partial^2_\lambda \Upsilon (T; \lambda)|_{\lambda=0}}{
\left(\partial_\lambda\Upsilon(T;\lambda)|_{\lambda=0}\right)^2}-1\right)^{1/2},
\label{gamma}
\end{eqnarray}
which implies that the effective broadness of $p(S_T(f)=S)$ is entirely defined
by the first two moments of the random variable $\tau_T$ in (\ref{tau}).
Specifically, it is independent of $D_0$ and $f$ when $f$ is only large enough.

(iv) The behaviour of the left tail of $p(S_T(f) = S)$ can be assessed in the
following way. Note that the product of the two Bessel functions in
(\ref{distlargef3}) can be represented as a power series with an infinite radius
of convergence (see (\ref{Bes}) in \ref{C}). Inserting the expansion in
(\ref{Bes}) in (\ref{distlargef3}) and integrating over $z$ we find
\begin{equation}
\fl \quad P(A)=\frac{2}{\sqrt{3}}\sum_{n=0}^{\infty}\frac{(-1)^n}{n!}\left(\frac{\sqrt{3}
+1}{\sqrt{6}}\right)^{2 n} {_2F_1}\left(-n,-n;1;\frac{1-\sqrt{3}/2}{1+\sqrt{3}/2}\right)
\left\langle \frac{T^{n+1}}{\tau_T^{n+1}} \right \rangle_{\Psi} A^n,
\label{Bess}
\end{equation}
if the inverse moments of the variable $\tau_T$ exist (and do not grow too fast
with $n$). Therefore, the PDF $P(A)$ is an analytic function of $A$ in the vicinity
of $A=0$, with
\begin{equation}
\label{zero}
P(0)=\frac{2}{\sqrt{3}}\left\langle\frac{T}{\tau_T}\right\rangle_{\Psi}.
\end{equation}
We note that below we will encounter both situations when $P(A)$ is analytic and
when it is not. In the latter situation we will show that $p(S_T(f)=S)$ diverges
as $S\to0$, which can be already inferred from (\ref{zero}).

\section{Diffusivity modelled as squared Ornstein-Uhlenbeck process}
\label{sec3}

In this and the following sections we apply the above general theory to several
random diffusivity models. According to our main results (\ref{mainfinf3}) and
(\ref{distlargef3}) one first needs to evaluate the MGF $\Upsilon(T;\lambda)$ of
the integrated diffusivity $\tau_T$ for a chosen diffusivity process $\Psi_t$. To
illustrate the quality of the theoretical predictions in the high-frequency limit
we also performed numerical simulations using a Python code.
The Euler integration scheme is used to compute (\ref{lan}), where $\Psi_t$ is
obtained by a numerical integration of the proper stochastic equation for each
case. The PSD is obtained by fast Fourier transform for each trajectory.
Starting from the single-trajectory power spectra the random amplitude $A$ is
calculated according to (\ref{eq:ST_A}).

Concretely when $\Psi_t$ in the diffusing-diffusivity model is defined as a stochastic
process satisfying some Langevin equation, the distribution of $A$ is determined
by (\ref{mainfinf3}) and (\ref{distlargef3}) through the MGF $\Upsilon(T;\lambda)$
of the integrated diffusivity $\tau_T$ that can be obtained by solving the
associated backward Fokker-Planck equation (see \cite{Grebenkov19} for details).
Here we consider the common example of squared Ornstein-Uhlenbeck process
and related models. We note that diffusing-diffusivity models are intimately
related to random-coefficient autoregressive processes \cite{jakub1}.

The Ornstein-Uhlenbeck process $Y_t$
\begin{equation}
\dot{Y}_t=-\tau_{\star}^{-1}Y_t+\sigma_\star\xi'_t
\end{equation}
is a stationary Gaussian process mean-reverting to zero at a time scale $\tau_
\star$ and driven by standard Gaussian white noise $\xi'_t$ with volatility
$\sigma_\star$. The process $\Psi_t=Y_t^2$ is one of the most common models of
diffusing diffusivity, which satisfies, due to the It\^o's formula,
\begin{equation}  
\label{eq:Feller}
\dot{\Psi}_t=\tau^{-1}(\bar{\Psi}-\Psi_t)+\sigma\sqrt{2\Psi_t}\xi'_t,
\end{equation}
where $\tau=\tau_\star/2$, $\sigma=\sqrt{2}\sigma_\star$, and $\bar{\Psi}=
\sigma_\star^2\tau_\star/2=\sigma^2\tau/2$. This model was extended in
\cite{Jain16,Jain16b,Chechkin:DD2} by considering $\Psi_t$ as the sum of $n$
independent squared Ornstein-Uhlenbeck processes, when (\ref{eq:Feller}) still
holds with $\bar{\Psi}=n\sigma^2\tau/2$. More generally, setting $\bar{\Psi}$
to be any positive constant, the Langevin equation (\ref{eq:Feller}) defines
the so-called Feller process \cite{Feller51}, also known as square root
process or the Cox-Ingersoll-Ross process \cite{Cox85}, also used in the Heston
model \cite{Heston93} . 
This process was used to model the diffusing-diffusivity in \cite{Lanoiselee18a,Lanoiselee18b}, 
see also the discussion in \cite{Chechkin:DD2}.

\begin{figure*}
\centering
\includegraphics[height=\textwidth,angle=270]{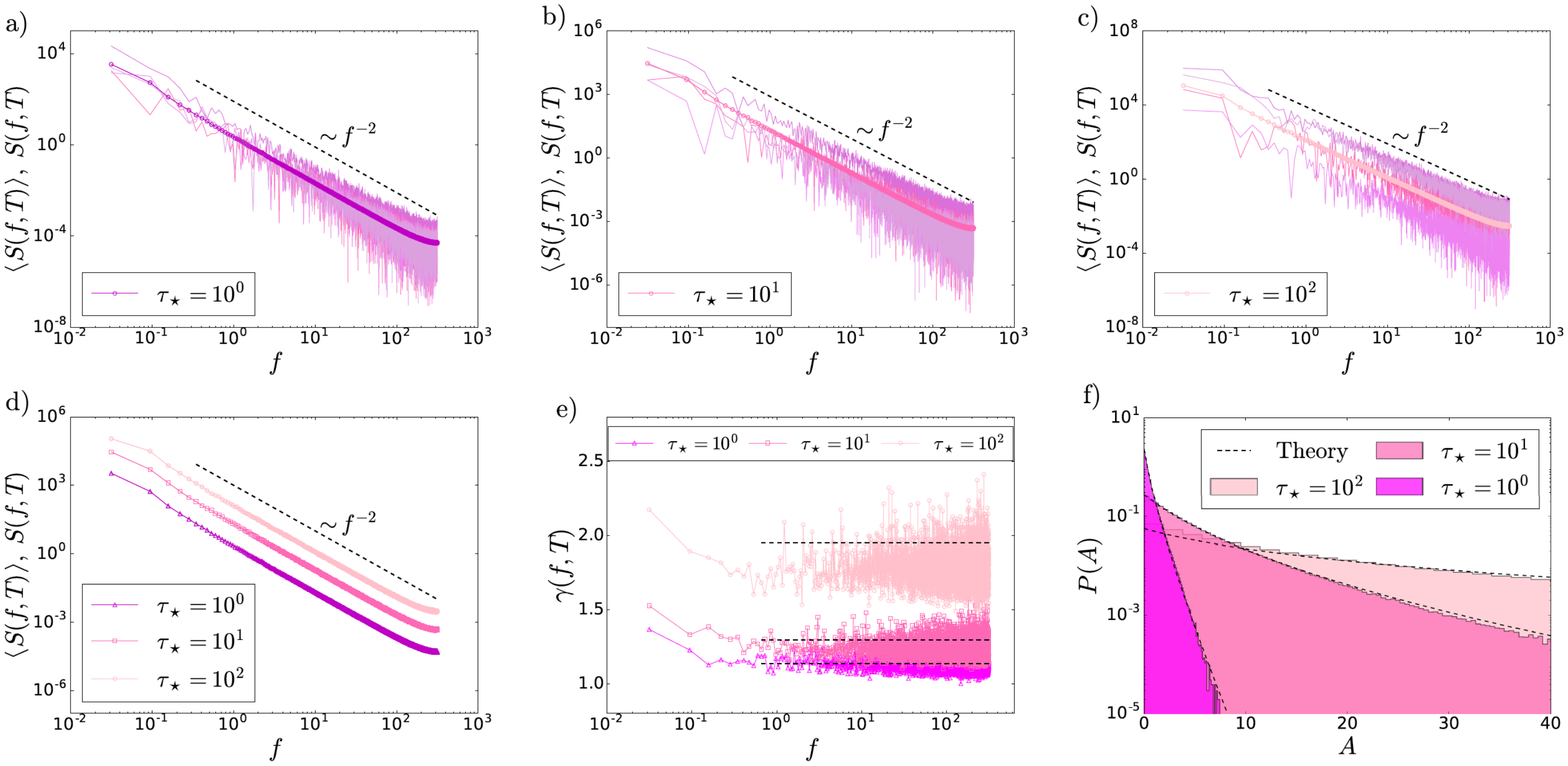}
\caption{Diffusing-diffusivity defined as the squared Ornstein-Uhlenbeck
process, for three different values of $\tau_\star$. Other parameters are
$T=10^2$ and $\sigma_\star=1$. Panels a)-c) show a few realisations of the
power spectra from individual trajectories and the average value for $\tau_
\star=10^0,10^1,10^2$, respectively. Panel d) shows a comparison of the
mean power spectrum. Panel e) shows the coefficient of variation, the
black dashed lines correspond to the theoretical result in (\ref{gam_ou}).
Panel f) shows the distribution of the random amplitude $A$, the black dashed
lines correspond to (\ref{distlargef3}) with the explicit expression of the
MGF from (\ref{upsilon1}).}
\label{fig6}
\end{figure*}

The MGF $\Upsilon(T;\lambda)$ for the integrated squared
Ornstein-Uhlenbeck process was first computed by Dankel \cite{Dankel} and employed
in \cite{Jain16,Jain16b,Chechkin:DD2}. Its computation for the Feller process in
(\ref{eq:Feller}) was presented in \cite{Lanoiselee18a},
\begin{equation}
\label{upsilon1}
\Upsilon(T;\lambda)=\left(\frac{4\omega e^{-(\omega-1)T/(2\tau)}}{(\omega+1)^2-
(\omega-1)^2 e^{-\omega T/\tau}}\right)^\nu, 
\end{equation}
where $\omega=\sqrt{1+4\sigma^2\tau^2\lambda}$ and $\nu=\bar{\Psi}/(\tau
\sigma^2)$. In particular, setting $\bar{\Psi}=\sigma^2\tau/2$ (and thus $\nu=
1/2$) one retrieves the MGF for the squared Ornstein-Uhlenbeck process. A
detailed discussion on the PDF of this model is presented in
\cite{Jain16,Jain16b,Chechkin:DD2,Lanoiselee18a,Lanoiselee18b}. Using the
explicit formulas for $\left \langle \overline{x_T^2}\right\rangle_Psi $ and $\left \langle \overline{x_T^4}\right\rangle_\Psi$
from \cite{Chechkin:DD2,Lanoiselee18a} we get from (\ref{gamma}) that
\begin{equation}
\gamma=\left[\frac{3}{4}\left(3+ \frac{6 \tau}{\nu T}\left(1-\frac{\tau}{T}
(1- e^{-T/\tau})\right) \right)-1 \right]^{1/2}.
\label{gam_ou}
\end{equation}
Moreover, as the second moment $\langle\overline{x_T^2}\rangle$ shows a linear
trend in time \cite{Chechkin:DD2,Lanoiselee18a}, no ageing of the PSD occurs,
as suggested in (\ref{muT}).

The PDF of $A$ is determined via (\ref{distlargef3}). Since an explicit
calculation of this integral is not straightforward we perform a numerical
integration. The results are shown in Fig.~\ref{fig6}, in which we observe
excellent agreement between the simulations and the theoretical results. The
$1/f^2$ scaling is recovered for any value of $\tau_\star$. The coefficient
of variation $\gamma$ converges to different values when we change $\tau_\star$,
according to (\ref{gam_ou}). Note that this result reflects the different
degrees of broadness of the PDF of the random amplitude $A$.
In particular for $\tau_\star\ll T$ we obtain a result that is very similar to
the one of Brownian motion, while for increasing $\tau_\star$ the PDF of the
random amplitude $A$ becomes increasingly broader.

\section{Diffusivity modelled as a jump process}
\label{l}

We divide the interval $(0,T)$ into $N$ equal subintervals of duration $\delta=
T/N$ and suppose that $\Psi_t$ is a jump process on these intervals, of the form
\begin{equation}
\Psi_t=\psi_k\mbox{ on }t\in\bigl([k-1]\delta,k\delta\bigr),\quad k=1,\ldots,N.
\end{equation}
Furthermore we stipulate that the $\psi_k$ are independent, identically distributed,
positive-definite random variables with PDF $\rho(\psi)$. In other words, we take
that $\Psi_t$ at each discrete time instant $(k-1)\delta$ attains a new random value, taken from the common distribution, and stays constant and equal to this value up to the next discrete instant $k\delta$. 
For a given realisation of the process
$\Psi_t$ we thus have
\begin{equation}
\tau_T=\int^T_0dt\Psi_t=\delta\sum_{k=1}^N\psi_k,
\label{mainJump}
\end{equation}
and hence
\begin{equation} 
\label{eq:Ups_jump}
\Upsilon(T;\lambda)
=\left(\int^{\infty}_0d\psi\rho(\psi)e^{-\lambda\delta\psi}\right)^{T/\delta}.
\end{equation}
Evaluating explicitly the derivatives $\partial_\lambda\Upsilon(T;\lambda)$ and
$\partial^2_\lambda\Upsilon (T;\lambda)$ at $\lambda=0$, we get 
\begin{equation}
\langle \tau_T \rangle =  T\,  {\mathbb E}\{\psi_k\}, 
\label{tauJump} 
\end{equation}
when the first moment ${\mathbb E}\{\psi_k\}$ exists. From this we infer $\left
\langle\overline{x_T^2}\right\rangle_\Psi$ and thus the respective ageing behaviour.
Moreover, the coefficient of variation becomes
\begin{equation}
\gamma=\left[\frac{9}{4}\left(1-\frac{\delta}{T}+\frac{\delta}{T}\frac{{\mathbb E}
\{\psi_k^2\}}{{\mathbb E}\{\psi_k\}^2}\right)-1 \right]^{1/2},
\label{gamma_jump}
\end{equation}
when the first two moments ${\mathbb E}\{\psi_k\}$ and ${\mathbb E}\{\psi_k^2\}$
exist.

Modelling the diffusivity as a jump process can be seen as a way to describe
the model in section \ref{sec3} through a different parametrisation. Indeed,
we define a time scale, which is given by the duration $\delta$ of each step
interval, and we then introduce a random variability of the diffusivity
from one interval to the next. These diffusivity fluctuations are chosen
according to the PDF $\rho(\psi)$. Of course, the main difference comes form
the fact that in this model we do not have any correlation between successive
diffusivities. In what follows we analyse two examples in detail. In the
first one we select a Gamma distribution for $\rho(\psi)$, in analogy with
the diffusing-diffusivity model in section \ref{sec3}, where the diffusion
coefficient shows a Gamma distribution as well. In the second example we select
a L\'evy-Smirnov distribution for $\rho(\psi)$. This allows us to model a
system in which a high probability of having small values of the diffusivity
is combined with the presence of few outliers, which can be related, for
instance, to values of the diffusivity at boundaries of the system.

\begin{figure*}
\centering
\includegraphics[height=\textwidth,angle=270]{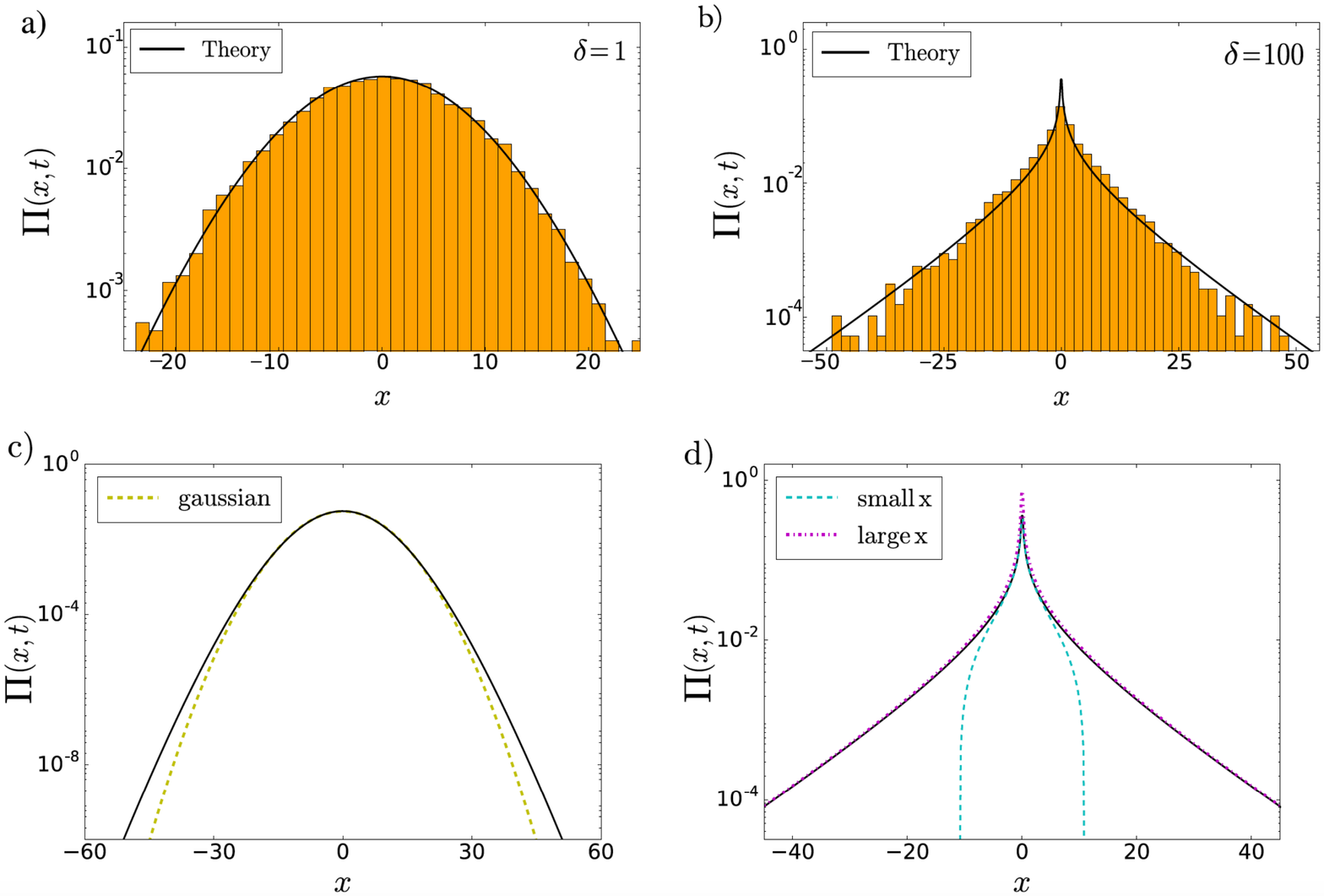}
\caption{Position-PDF at a fixed time $t=50$ for diffusivity modelled as a jump process
with Gamma distribution ($\psi_0=1$ and $\nu=0.5$). Panels a) and c) correspond
to $\delta=1$, and panels b) and d) to $\delta=100$. Panels a) and b) show a
comparison between the numerical and the analytic result in (\ref{pdf_2a}) (black
dashed lines). Panel c) shows a comparison between the analytic result (\ref{pdf_2a})
for $\delta=1$ and its Gaussian approximation (\ref{47}). Panel d) compares between
the analytical result (\ref{pdf_2a}) for $\delta=100$ and its asymptotic behaviours
in (\ref{39}) and (\ref{40}).}
\label{fig_prop1}
\end{figure*}

\subsection{Example I: Gamma distribution}

\begin{figure*}
\centering
\includegraphics[height=\textwidth,angle=270]{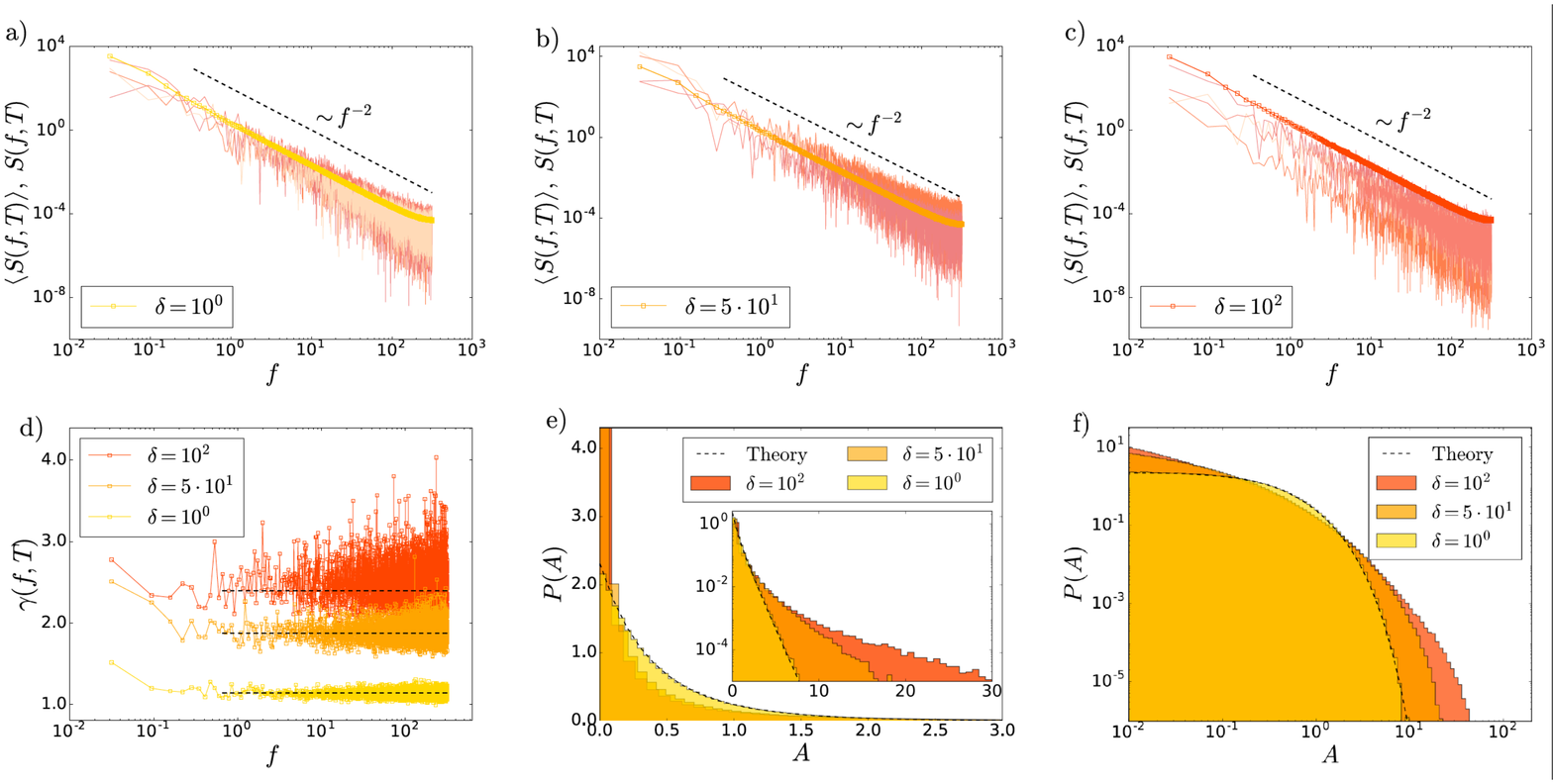}
\caption{Diffusivity modelled as a jump process with Gamma distribution ($\psi
_0=1$ and $\nu=0.5$), for varying $\delta$ and with trajectory length $T=10^2$.
Panels a)-c) show a few realisations of the power spectra from individual
trajectories and the mean value for $\delta=10^2$, $5\times10^1$, and $10^0$,
respectively. Panel d) shows the coefficient of variation for three values of
$\delta$---the black dashed lines represent the theoretical result (\ref{gamma_jump}).
Panels e)-f) depict the distribution of the random amplitude $A$ in linear, semi-log
(inset) and log-log scale. Here the black dashed line corresponds to the theoretical
result (\ref{distlargef4a}).}
\label{fig2}
\end{figure*}

First, we consider the Gamma distribution,
\begin{equation}  
\label{gamma_dist}
\rho(\psi)=\frac{\psi^{\nu-1}}{\Gamma(\nu)\psi_0^{\nu}}\exp\left(-\psi/\psi_0
\right)
\end{equation}
with the shape parameter $\nu>0$ and the scale parameter $\psi_0>0$.  From
(\ref{eq:Ups_jump}), we deduce
\begin{equation}
\Upsilon(T;\lambda)=\left(1+\lambda\delta\psi_0\right)^{-\nu T/\delta}.
\label{mgf_gam}
\end{equation}

\subsection*{Position-PDF ${\bf \Pi}(x,t)$}

A direct calculation of the PDF for this model can be performed. Starting
from (\ref{pdf_DD}) and recalling that $\Phi_w=\Upsilon(T;D_0w^2)$, we get
\begin{equation}
\label{pdf_2a}
{\bf \Pi}(x,t)={\cal N}_t\left(\frac{ |x|}{2\sqrt{D_0\delta\psi_0}}\right)^{\frac{\nu
t}{\delta}-\frac{1}{2}}K_{\frac{1}{2}-\frac{\nu t}{\delta}}\left(\frac{|x|}{ \sqrt{D_0\delta\psi_0}}\right),
\end{equation}
where the normalisation coefficient is
\begin{equation}
{\cal N}_t=\sqrt{\frac{2}{\pi}}\Big/\left(\left(D_0\delta\psi_0\right)^{3/4}
\Gamma\left(\frac{\nu t}{\delta}\right)\right).
\end{equation}
With the properties
\begin{equation}
z^{\nu}K_{-\nu}(z)\sim2^{\nu-1}\Gamma(\nu)-\frac{2^{\nu-3}\Gamma(\nu)}{\nu-1}z^2
\end{equation}
for $|z|\to0$ and $\nu>1$, as well as
\begin{equation}
K_{-\nu}\sim\sqrt{\frac{\pi}{2z}}e^{-z}
\label{K_long}
\end{equation}
for $|z|\to\infty$, the asymptotic behaviours of the PDF are given by
\begin{equation}
{\bf \Pi}(x,t)\sim {\cal N}_t \, 2^{\frac{\nu t}{\delta}-\frac{3}{2}} \Gamma
\left(\frac{\nu t}{\delta}-\frac{1}{2}\right)\left[1-\frac{x^2}{4\left(\frac{\nu t }{ \delta}
-\frac{3}{2}\right)D_0\delta\psi_0}\right]
\label{39}
\end{equation}
for $|x|\to0$ and $\nu t>3\delta/2$, as well as
\begin{equation}
{\bf \Pi}(x,t)\sim\frac{2}{\sqrt{D_0\delta\psi_0}\Gamma\left(\frac{\nu t}{\delta}\right)}\left(\frac{|x|}{2 \sqrt{D_0\delta\psi_0}}\right)^{\frac{\nu t}{\delta}-1} \exp\left(-\frac{|x|}{\sqrt{D_0 \delta \psi_0}}\right)
\label{40}
\end{equation}
for $|x|\to\infty$.

The functional behaviour of the PDF ${\bf \Pi}(x,t)$ is shown in
Fig. \ref{fig_prop1}. We see that by changing $\delta$ we can observe different
shapes of ${\bf \Pi}(x,t)$. When $\delta=1$ (panels a) and c)) the Gaussian approximation
(\ref{47}) already provides a good estimate of the PDF over a wide range. We
start observing discrepancies only far out in the tails, for values which can hardly
be reached with real data. When $\delta=100$ (panels b) and d)), in contrast, the
exponential tails are distinct. The behaviours at small and large $x$ are well
described by the asymptotic expansions in (\ref{39}) and (\ref{40}).
Note that the value of $\delta$ in here plays a role similar to the correlation
time $\tau_\star$ in the diffusing-diffusivity model defined in section
\ref{sec3}. The only difference is that by changing $\tau_\star$ in the
model above we also change the average diffusivity while, in this case,
changes in the value of $\delta$ do not affect the average diffusivity,
which is fixed once we choose the jumps PDF in (\ref{gamma_dist}).

\subsection*{Amplitude-PDF $P(A)$}

The MGF of the amplitude $A$ of the jump process-diffusivity model is given by
\begin{equation}
\label{mainfinf4}
\Phi_{\lambda}=\frac{2}{\sqrt{3}}\int^{\infty}_0dp\frac{\displaystyle \exp\left(-4p/3\right) 
I_0(2 p/3)}{\displaystyle \left(1+p\psi_0\lambda\delta/T\right)^{\nu T/\delta}},
\end{equation}
so that
\begin{equation}
\fl \qquad P(A)=\frac{2}{\sqrt{3}}\int^{\infty}_0dz\frac{\displaystyle J_0\left(\left(1+1/\sqrt{3}\right) 
\sqrt{2zA}\right)}{\displaystyle \left(1+z\psi_0\delta/T\right)^{\nu T/\delta}} J_0\left(\left(1-1/\sqrt{3}\right)\sqrt{2zA}\right).
\label{distlargef4}
\end{equation}
In particular one has ${\mathbb E}\{\psi_k\}=\nu\Psi_0$ and ${\mathbb E}\{\psi_k^2
\}=\Psi_0^2\nu(\nu+1)$, thus
from (\ref{tauJump}) we readily obtain $\left\langle\overline{x_T^2}\right\rangle
_\Psi\simeq T$, demonstrating that in this process no ageing behaviour is displayed.
Moreover, from (\ref{gamma_jump}) we get
\begin{equation}
\gamma=\left[\frac{9}{4}\left(1+\frac{\delta}{\nu T}\right)-1\right]^{1/2}.
\label{gam_gam}
\end{equation}
In the limit $\delta\to0$ and $N\to\infty$, with $\delta N=T$ fixed, we have
\begin{equation}
\Upsilon(T;\lambda)\sim\exp\left(-\nu\psi_0T\lambda\right).
\label{short_delta}
\end{equation}
Hence, 
\begin{eqnarray}
\nonumber
\Phi_{\lambda}&=&\frac{2}{\sqrt{3}}\int^{\infty}_0dp\exp\left(-\left(\frac{4}{3}
+\nu\psi_0\lambda\right)p\right)I_0\left(\frac{2 p}{3}\right)\\
&=&\left(1+2\nu\psi_0\lambda+\frac{\displaystyle 4}{\displaystyle 3}(\nu\psi_0\lambda)^2\right)^{-1/2} 
\label{mainfinf4a}
\end{eqnarray}
and
\begin{eqnarray}
\nonumber
\fl \qquad P(A)&=&\frac{2}{\sqrt{3}}\int^{\infty}_0dzJ_0\left(\left(1+1/\sqrt{3}\right)
\sqrt{2 z A}\right) J_0\left(\left(1-1/\sqrt{3}\right)\sqrt{2zA}\right)e^{-\nu\psi_0z}\\
\fl &=&\frac{2}{\sqrt{3}\nu\psi_0}\exp\left(-\frac{4A}{3\nu\psi_0}\right)I_0\left(
\frac{2 A}{3\nu\psi_0}\right).
\label{distlargef4a}
\end{eqnarray}
This means that we have essentially the same behaviour as for standard
one-dimensional Brownian motion, however, with renormalised coefficients
(compare with the result in \cite{we1}),
in agreement also with what we obtained for the diffusing-diffusivity model
in section \ref{sec3}, when selecting $\tau_\star\ll T$.
Indeed, if we use (\ref{short_delta}) and recall that $\Phi_w=\Upsilon(T;D_0
w^2)$, we readily obtain
\begin{equation}
{\bf \Pi}(x,t)\sim\frac{1}{2\sqrt{\pi\nu\psi_0D_0t}}\exp\left(-\frac{x^2}{4\nu\psi_0
D_0t}\right).
\label{47}
\end{equation}

In Fig. \ref{fig2} we show a direct comparison between the numerical and theoretical
results for the Gamma distribution with $\psi_0=1$ and $\nu=0.5$. We observe that
the average value of the power spectrum is not affected by the value of $\delta$.
Nevertheless, when we plot some sample single-trajectory power spectra we notice
a larger amplitude scatter for larger values of $\delta$. This may be clearly
seen in the distribution of the random variable $A$, which is broader for larger
values of $\delta$, and consequently in the different limiting values of the
coefficient of variation. Thus, the fluctuations are sensitive to different
parameters of the distribution (\ref{gamma_dist}), while the mean behaviour is
not.

\subsection{Example II: L\'evy-Smirnov distribution}

In our second example we consider the L\'evy-Smirnov distribution
\begin{equation}  
\label{LS_dist}
\rho(\psi)=\sqrt{\frac{\displaystyle \psi_0}{\displaystyle \pi}}\frac{\displaystyle \exp\left(-\psi_0/\psi\right)}{\displaystyle 
\psi^{3/2}},
\end{equation}
for which Eq. (\ref{eq:Ups_jump}) yields
\begin{equation}
\Upsilon(T;\lambda)=\exp\left(-2T\sqrt{\psi_0\lambda/\delta}\right).
\label{mgf_levy}
\end{equation}
Note that in this case ${\mathbb E}\{\psi_k\}$ and ${\mathbb E}\{\psi_k^2\}$
are not defined, such that $\left\langle\overline{x_T^2}\right\rangle_\Psi$
does not exist either. This suggests that a clear ageing behaviour cannot be
defined and that fluctuations are what dominates the system.

\subsection*{Position-PDF ${\bf \Pi}(x,t)$}

As a consequence, we obtain the following analytical expression for the PDF,
\begin{equation}
{\bf \Pi}(x,t)=\frac{2t\sqrt{D_0\psi_0/\delta}}{\pi}\frac{1}{4t^2D_0\psi_0/\delta
+x^2},
\label{pdf_2b}
\end{equation}
where we recognise the power-law behaviour, that is already built into relation
(\ref{LS_dist}). Note that expression (\ref{pdf_2b}) represents the Cauchy
distribution, whose median grows with time $t$.

\begin{figure*}
\centering
\includegraphics[height=\textwidth,angle=270]{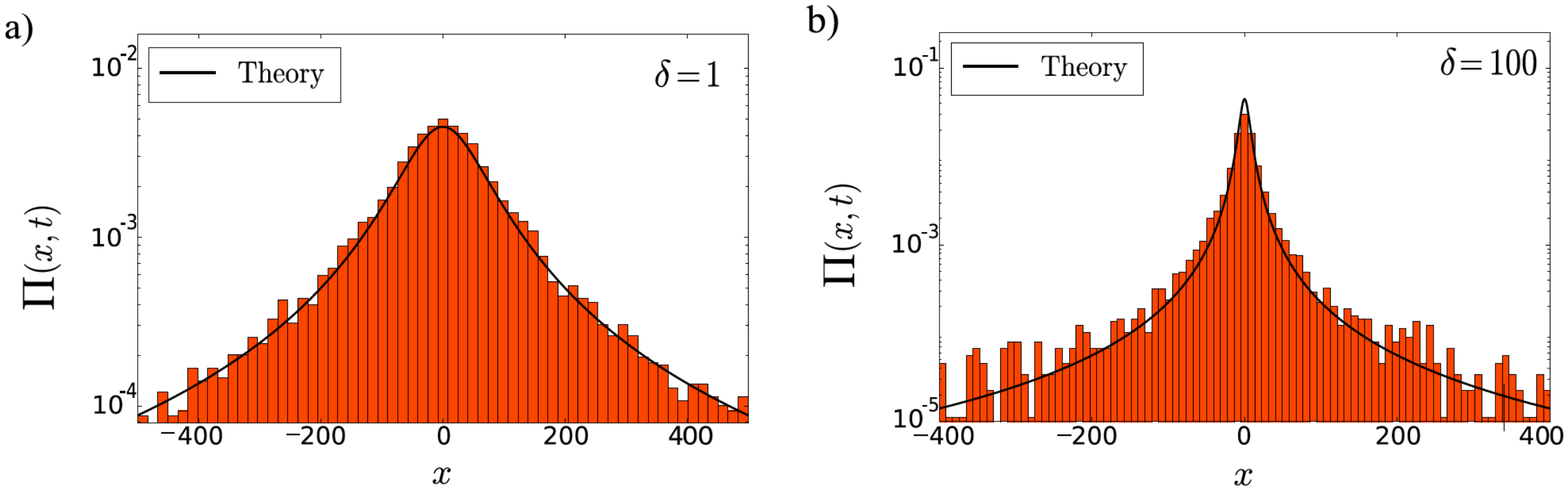}
\caption{Position-PDF at fixed time $t=50$ for diffusivity modelled as a jump process
with L\'evy-Smirnov distribution, $\Psi_0=0.5$, for a) $\delta=1$ and b)
$\delta=100$. The black dashed line represents the analytical result
(\ref{pdf_2b}).}
\label{fig_prop2}
\end{figure*}

The PDF is shown for two different values of $\delta$ in Fig.~\ref{fig_prop2}.
We observe that, differently from the case with the Gamma distribution above,
we do not see significant changes in the shape of the distribution when varying
$\delta$. For both cases, $\delta=1$ and $\delta=100$, the power-law behaviour
(\ref{pdf_2b}) is readily discernible.

\subsection*{Amplitude-PDF $P(A)$}

\begin{figure*}
\centering
\includegraphics[height=\textwidth,angle=270]{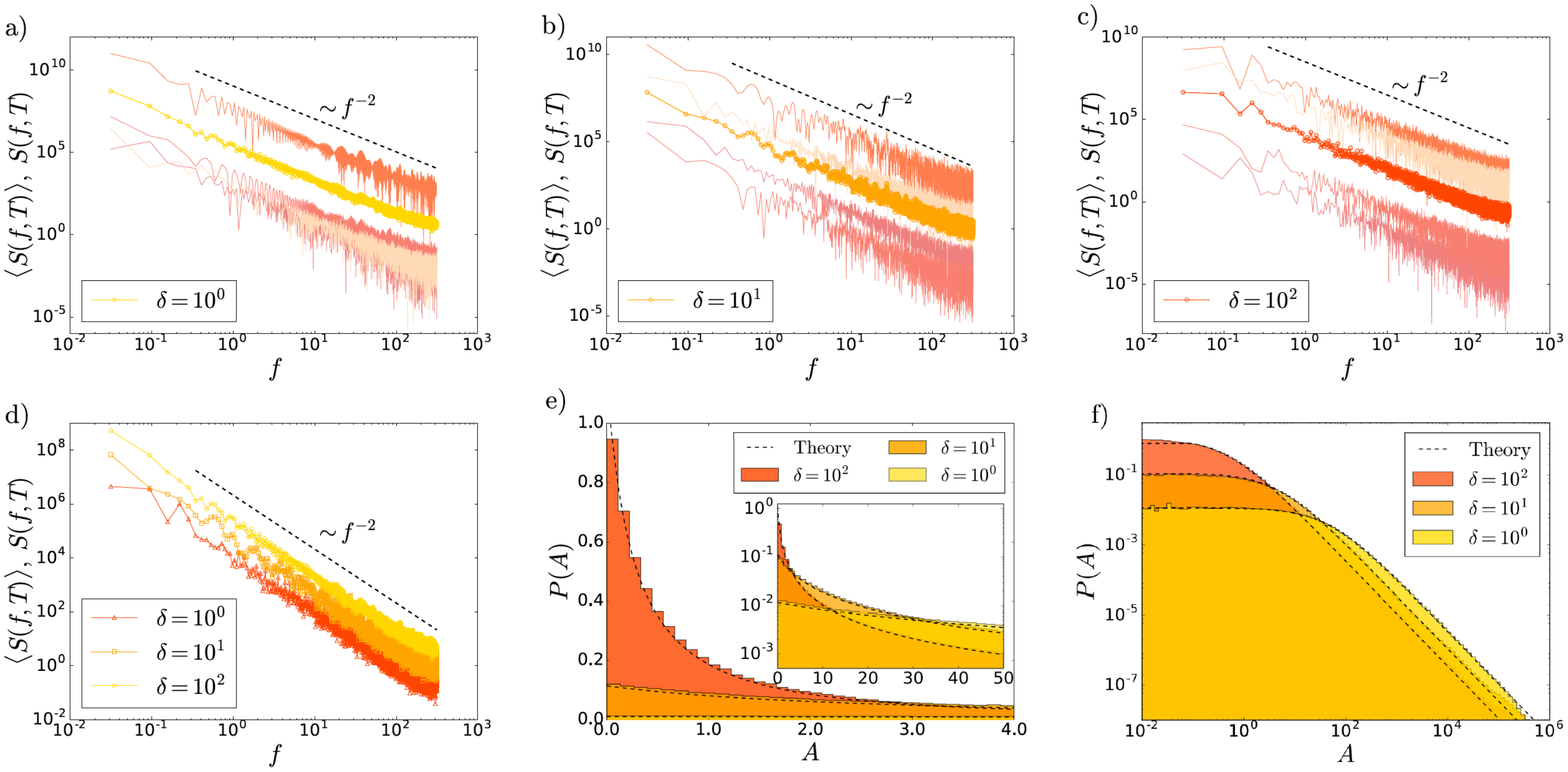}
\caption{Diffusivity modelled as a jump process with L\'evy-Smirnov distribution
with $\psi_0=0.5$, for varying $\delta$ and with trajectory length $T=10^2$.
Panels a)-c) show a few realisations of the power spectra from individual
trajectories and the mean value for $\delta=10^2$, $10^1$, and $10^0$,
respectively. Panel d) compares the mean values of the power spectrum. Panels
e)-f) show the distribution of the random amplitude $A$ in linear, semi-log
(inset) and log-log scale---the black dashed line corresponds to the analytical
result (\ref{distlargef5}).}
\label{fig5}
\end{figure*}

The MGF for the random amplitude $A$ reads
\begin{equation}
\label{mainfinf5}
\Phi_{\lambda}=\frac{2}{\sqrt{3}}\int^{\infty}_0dp\exp\left(-\frac{4}{3}p  
-2\sqrt{p\psi_0\lambda T/\delta}\right)I_0\left(\frac{2p}{3}\right)  
\end{equation}
and
\begin{eqnarray}
\nonumber
P(A)&=&\frac{2}{\sqrt{3}}\int^{\infty}_0dzJ_0\left(\left(1+1/\sqrt{3}\right)
\sqrt{2zA}\right)\\
\nonumber
&&\times J_0\left(\left(1-1/\sqrt{3}\right)\sqrt{2 z A}\right)
\exp\left(-2\sqrt{z\psi_0T/\delta}\right)\\
&=&\frac{\delta}{\sqrt{3}\psi_0T}\frac{1}{(1+\xi)^{3/2}}
\,_2F_1\left(\frac{3}{4},\frac{5}{4};1;\frac{\xi^2}{4(1+\xi)^2}\right),
\label{distlargef5}
\end{eqnarray} 
with $\xi=(4A\delta)/(3\psi_0T)$. Note that in the limit $A\to\infty$, the
leading behaviour of $P(A)$ follows
\begin{equation}
P(A)\sim\frac{1}{A^{3/2}}.
\end{equation}
Thus, the PDF $P(A)$ inherits the property of diverging moments from the
parental L\'evy-Smirnov distribution.

Figure \ref{fig5} summarises the properties of the PSD for the jump process
with L\'evy-Smirnov distribution ($\psi_0=0.5$). We observe that, despite the
fat-tailed PDF in (\ref{pdf_2b}) we still observe the universal $1/f^2$
scaling of the PDF. Concurrently, the PDF of the random amplitude $A$ features
the power-law behaviour according to (\ref{distlargef5}). Note that the
non-existence of the moments of $P(A)$ generates a pronounced scatter in the
amplitude of the average power spectrum.

\section{Diffusivity modelled as a functional of Brownian motion}
\label{m}

We now focus on the case when $\Psi_t$ is a genuine "diffusing-diffusivity" in
the sense that it is subordinated to Brownian motion $B_t$ starting at the
origin at $t=0$, with zero mean and covariance function
\begin{equation}
\left\langle B_tB_{t'}\right\rangle=2D_B\min\{t,t'\}.
\end{equation}
We here choose $\Psi_t=V[B_t]$, where $V$ is some prescribed, positive-definite
function. Note that random variables of the form $\int^T_0dtV[B_t]$ appear across
many disciplines, including probability theory, statistical analysis, computer
science, mathematical finance and physics. Starting from earlier works \cite{0,5,5a,6,7,8},
much effort has been invested in the analysis of the PDF and the
corresponding Laplace transforms of these processes. A large body of exact
results has been obtained within the last seven decades (see, e.g., \cite{1,2,3,
4,4a,Borodin} and further references therein). In the following, we consider three particular examples of $V[B_t]$, for which we can carry out exact calculations and obtain insightful results.

\subsection{Example I: $\Psi_t = \Theta(B_t)$}

First, we choose the cut-off Brownian motion
\begin{equation}
\Psi_t=\Theta(B_t),
\end{equation}
where $\Theta(x)$ is the Heaviside theta function. The process $x_t$
exhibits standard diffusive motion, once $B_t>0$, and pauses, remaining at
the position it has reached when $B_t$ goes to negative values. The random
variable $\int^T_0dt\Psi_t$ defines the time spent by a Brownian trajectory,
starting at the origin, on the positive real line within the time interval
$(0,T)$. The time intervals between any two "diffusion tours", as well as
their duration, are random variables with a broad distribution.

This example is of particular interest as it represents an alternative to
other standard models describing waiting times and trapping events. One
could think, for instance, of the comb model, where a particle, while
performing standard Brownian motion along one direction, gets stuck for
a random time in branches perpendicular to the direction of the diffusive
motion \cite{comb1,comb2}.

The MSD of the process $x_t$, as one can straightforwardly check, is just
\begin{equation}
\left\langle\overline{x_t^2}\right\rangle_{\Psi}=D_0t,
\end{equation} 
that is, a standard diffusion law in which the diffusion coefficient is reduced
by the factor $1/2$.
This means that no ageing behaviour is observed.
For higher order moments one expects, of course, significant
departures from the standard diffusive behaviour.

\begin{figure*}
\centering
\includegraphics[height=\textwidth,angle=270]{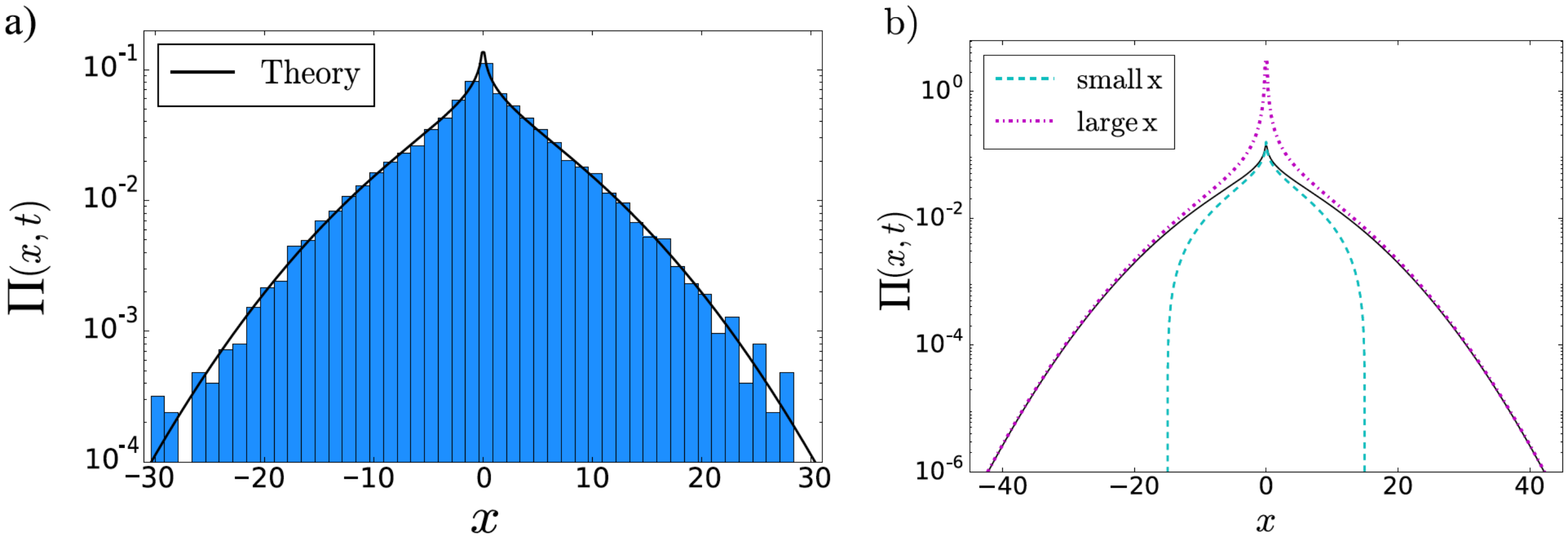}
\caption{Position-PDF at fixed time $t=50$ for diffusivity modelled as Heaviside
function of Brownian motion, with $D_B=1$. Panel a) compares the numerical results
with the analytical expression (\ref{pdf_3b}) (black solid line). Panel b) shows a
comparison between the analytical result and its asymptotic behaviours (\ref{74})
and (\ref{75}).}
\label{fig_prop8}
\end{figure*}

\subsection*{Position-PDF ${\bf \Pi}(x,t)$}

The MGF of the random variable $\tau_T=\int^T_0dt\Theta(B_t)$, which has a
bounded support on $(0,T)$, has been first derived by Kac \cite{6}, and
Erd\"os and Kac \cite{7}. Rewriting their result in our notation we have
\begin{eqnarray}
\nonumber
\Upsilon(T;\lambda)&=&\Bigg\langle\exp\left(-\lambda\int^T_0dt\Theta(B_t)
\right)\Bigg\rangle_{\Psi}\\
&=&e^{-\lambda T/2}I_0\left(\frac{\lambda T}{2}\right).
\label{t}
\end{eqnarray} 
Note that the inverse Laplace transform of this expression produces the
celebrated L{\'e}vy arcsine law \cite{arcsine}. With result (\ref{t}), the
desired PDF is given by
\begin{eqnarray}
\nonumber
{\bf \Pi}(x,t)&=&\frac{1}{\pi}\int^{\infty}_0dw\cos(wx)\exp\left(-\frac{D_0t}{2}
w^2\right) I_0\left(\frac{D_0 t}{2} w^2\right)\\
&=&\frac{\exp\left(-x^2/(8D_0t)\right)}{2\sqrt{\pi^3D_0t}}K_0\left(\frac{
x^2}{8D_0t}\right). 
\label{pdf_3b}
\end{eqnarray}
Recalling that $K_0(z)\sim-\ln(z/2)-\gamma_\mathrm{EM}$ for $|z|\to0$, where
$\gamma_\mathrm{EM}$ is the Euler-Mascheroni constant, for small $x$ we have
\begin{equation}
{\bf \Pi}(x,t)\sim\frac{-\ln\left(\frac{x^2}{8D_0t}\right)-\gamma_\mathrm{EM}}{
2\sqrt{\pi^3D_0t}},\quad|x|\to0.
\label{74}
\end{equation}
For large $x$ we use (\ref{K_long}) and obtain the asymptotic behaviour for the PDF,
\begin{equation}
{\bf \Pi}(x,t)\sim\frac{1}{\pi|x|}\exp\left(-\frac{x^2}{4D_0t}\right),\quad|x|\to\infty. 
\label{75}
\end{equation}

The PDF for this process is shown in Fig.~\ref{fig_prop8}. We see that the
central part of the PDF is strongly non-Gaussian, while the tails are
Gaussian, in agreement with the asymptotic behaviours (\ref{74}) and (\ref{75}).

\subsection*{Amplitude-PDF $P(A)$}

Inserting expression (\ref{t}) into (\ref{mainfinf3}) and performing the
integration over $z$, we arrive at the following, remarkably compact
expression for the MGF,
\begin{equation}
\label{step}
\Phi_{\lambda}=\frac{2\sqrt{2}}{\pi\sqrt{2+3\lambda}}{\bf K}\left(\frac{2
\lambda}{2+3\lambda}\right)
\end{equation}
where ${\bf K}(x)$ is the complete elliptic integral of the first kind,
\begin{equation}
\label{ellip}
{\bf K}(x)=\int^{\pi/2}_0\frac{d\phi}{\sqrt{1-x\sin^2(\phi)}}.
\end{equation}
Note that the high-$f$ asymptotic form in (\ref{step}) is independent
of the observation time $T$.

To proceed we take advantage of the definition of the complete elliptic integral
and perform the inverse Laplace transform of (\ref{step}). After
some formal manipulations this yields the following expression for the PDF,
\begin{equation}
\fl \qquad P\left(A\right)=2\sqrt{\frac{2}{3\pi^3A}}\int^{\pi/2}_0\frac{\displaystyle d\phi}{\displaystyle \sqrt{1-
\frac{2}{3}\sin^2(\phi)}}\exp\left(-\frac{\displaystyle 2A}{\displaystyle 3\left(1-\frac{2}{3}\sin^2(\phi)\right)}\right).
\label{PB}
\end{equation}
Multiplying both sides of the latter equation by $A^n$ and integrating over $A$
from $0$ to $\infty$, we get the following simple expression for the moments of
the random amplitude $A$ of an arbitrary order,
\begin{equation}
\label{momm}
{\mathbb E}\{A^n\}=\frac{\displaystyle \Gamma\left(n+\frac{1}{2}\right)}{\displaystyle \sqrt{\pi}}\left(
\frac{\displaystyle 3}{\displaystyle 2}\right)^n \,_2F_1\left(-n,\frac{1}{2};1;\frac{2}{3}\right).
\end{equation}
Then, from (\ref{gamma}), we readily get the coefficient of variation,
$\gamma=\sqrt{19/8}$. 

\begin{figure*}
\centering
\includegraphics[height=\textwidth,angle=270]{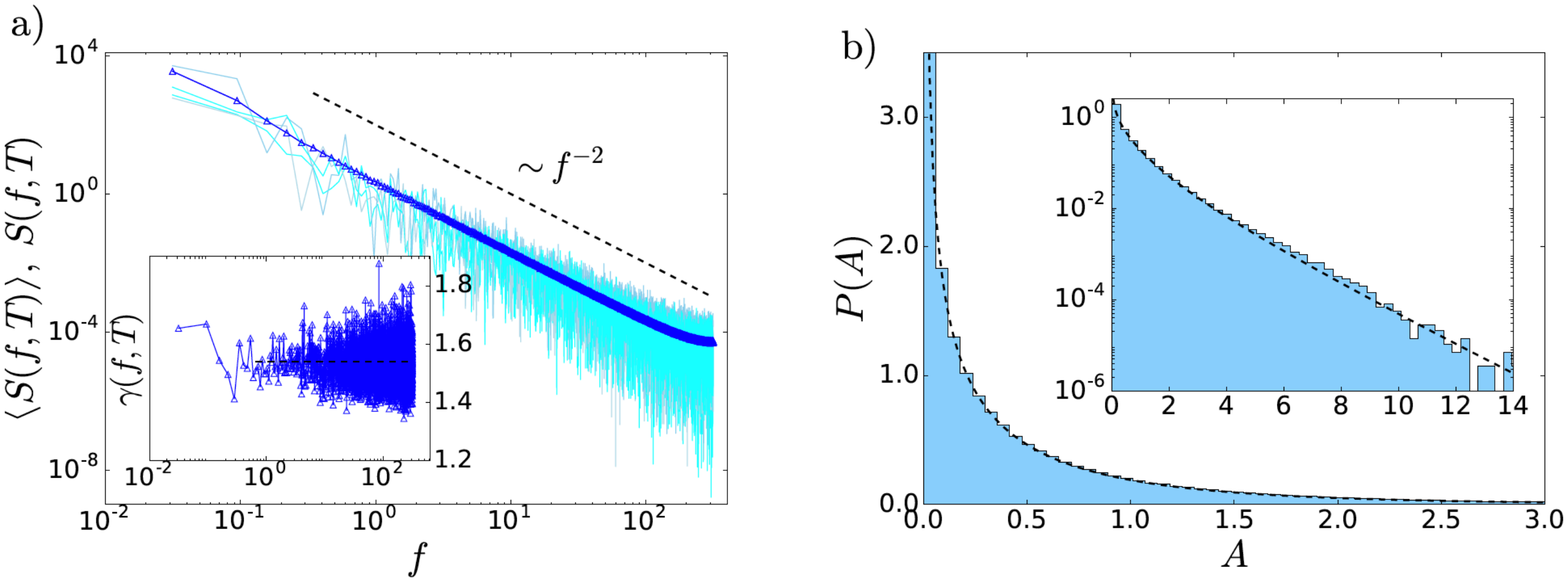}
\caption{Diffusivity modelled as Heaviside theta function of Brownian motion,
with $D_B=1$ and trajectory length $T=10^2$. Panel a) shows the mean power
spectrum along with a few realisations of power spectra from individual
trajectories. In the inset the coefficient of variation is shown, the black
dashed line indicates the theoretical value $\sqrt{19/8}\approx1.54$. Panel b) shows
the distribution of the random amplitude $A$. Here the black dashed line
corresponds to the numerical evaluation of (\ref{PB}).}
\label{fig8}
\end{figure*}

Note that the integrals $\int^{\infty}_0d\lambda\lambda^n\Upsilon(T;\lambda)$
diverge for any $n >0$, which means that $\tau_T$ does not have negative moments.
One therefore expects that $P(A)$ is a non-analytic, diverging function in the
limit $A\to0$. The small-$A$ asymptotic behaviour of $P(A)$ can be deduced directly from (\ref{PB}). Expanding the exponential function in the integral into the Taylor series in powers of $A$ and expressing the emerging generalised
elliptic integrals via their representations in terms of the toroidal
functions $P_{n-1/2}\left(\cosh(\eta)\right)$ (see (\ref{tor}) in \ref{C}), we get
\begin{equation}
\label{smallS}
P\left(A\right)=\sqrt{\frac{2}{\sqrt{3}\pi A}}\sum_{n=0}^{\infty}\frac{(-1)^n}{n!} 
\left(\frac{2}{\sqrt{3}}\right)^nP_{n-1/2}\left(\frac{2}{\sqrt{3}}\right)A^n.
\end{equation} 
For the opposite limit $A \to \infty$, we conveniently rewrite
Eq. (\ref{distlargef5}) in the form
\begin{eqnarray}
\nonumber
\fl \qquad P(A)&=&2\sqrt{\frac{2}{3\pi^3A}}\exp\left(-\frac{2A}{3}\right)\int^{\pi/2}_0
\frac{d\phi}{\sqrt{\left(1-\frac{2}{3}\sin^2(\phi)\right)}}
\exp\left(-\frac{2 A}{3}\frac{\frac{2}{3}\sin^2(\phi)}{1-\frac{2}{3}
\sin^2(\phi)}\right)\\
\nonumber
\fl &=&2\sqrt{\frac{2}{3\pi^3A}}\exp\left(-\frac{2A}{3}\right)\sum_{n=0}^{\infty}
\left(\int^{\pi/2}_0\sin^{2n}(\phi)\right)
\left(\frac{2}{3}\right)^n{\rm L}^{(-1/2)}_n\left(\frac{2A}{3}\right)\\
\fl &=&\frac{2}{3\pi}\sqrt{\frac{3}{2A}}\exp\left(-\frac{2A}{3}\right)\sum_{n=0}^{
\infty}\frac{\Gamma(n+1/2)}{n!}
\left(\frac{2}{3}\right)^n{\rm L}^{(-1/2)}_n\left(\frac{2A}{3}\right),
\end{eqnarray}
where ${\rm L}^{(-1/2)}_n(x)$ are associated Laguerre polynomials. We focus next
on the asymptotic behaviour of the function
\begin{equation}
g(u)=u^{-1/2}\sum_{n=0}^{\infty}\frac{\Gamma(n+1/2)}{n!}\left(\frac{2}{3}\right)^n
{\rm L}^{(-1/2)}_n\left(u\right)
\end{equation}
in the limit $u\to\infty$. Performing a Laplace transform of $g(u)$ we readily get
\begin{eqnarray}
\nonumber
{\cal L}_s\left\{g(u)\right\}&=&\int^{\infty}_0du\exp\left(-su\right)g(u)\\
\nonumber
&=&\frac{1}{s^{1/2}}\sum_{n=0}^{\infty}\frac{\Gamma^2(n+1/2)}{(n!)^2}\left(\frac{2
(s-1)}{3s}\right)^n\\
&=&\frac{2}{s^{1/2}}{\bf K}\left(\frac{2(s-1)}{3s}\right),
\end{eqnarray}
where ${\bf K}$ is the complete elliptic integral defined in Eq. (\ref{ellip}).
In the limit $s\to0$ (corresponding to $A\to\infty$),
\begin{equation}
{\cal L}_s\left\{g(u)\right\}\sim\frac{2}{s^{1/2}}{\bf K}\left(-\frac{2}{3s}
\right)=2\int^{\pi/2}_0\frac{d\phi}{\sqrt{s+\frac{2}{3}\sin^2(\phi)}}.
\end{equation}
Inverting the Laplace transform and integrating over $\phi$, we find
\begin{equation}
g(u)\sim\frac{\pi}{\sqrt{u}}\exp\left(-\frac{u}{3}\right)I_0\left(\frac{u}{3}
\right)\to\sqrt{\frac{3}{2}}\frac{1}{u}.
\end{equation}
Thus, in the limit $A\to\infty$, the leading behaviour of the PDF $P(A)$ yields
in the form
\begin{equation}
P(A)\sim\frac{1}{\pi A}\sqrt{\frac{3}{2}}\exp\left(-\frac{2A}{3}\right).
\end{equation}
Figure \ref{fig8} summarises the numerical results for this case. Again we
observe excellent agreement with the theoretical results. 

\subsection{Example II: $\Psi_t=\exp(-B_t/a)$.}

As the second example, we link the process $\Psi_t$ to so-called geometric Brownian motion. Random variables of this form have been widely studied in the mathematical finance literature (see, e.g., \cite{1}). Within the latter domain, they emerge very naturally as representation of the solution of the celebrated
Black-Scholes equation. Their time-averaged counterpart is related to the
so-called Asian options \cite{51,61,71} (see also \cite{peters,deepak})
and also appears in different contexts in the analysis of transport phenomena
in disordered media (see, e.g., \cite{100,101,102,103,104,104a,104b}) as well as
characterises some features of the melting transition of heteropolymers \cite{105}.

In our notation, we set
\begin{equation}
\Psi_t=\exp\left(-\frac{B_t}{a}\right),
\end{equation} 
where $a$ is a parameter of unit length. In this case, $x_t$ exhibits an
anomalously strong superdiffusion such that
\begin{eqnarray} 
\nonumber
\left\langle\overline{x_t^2}\right\rangle&=&2D_0\int^t_0d\tau\langle\Psi_{\tau}
\rangle_{\Psi}=2D_0\int^t_0d\tau\exp\left(D_B\tau/a^2\right)\\
&=&\frac{2D_0a^2}{D_B}\left(\exp\left(D_B t/a^2\right)-1\right).
\end{eqnarray}
Note that when $D_B t/a^2\ll1$ we have $\left\langle\overline{x_t^2}\right\rangle
\sim2D_0t$. These results for the MSD demonstrate that for this model, in general,
we observe ageing behaviour, though the latter may be hidden while analysing very
short trajectories.

\subsection*{Position-PDF ${\bf \Pi}(x,t)$}

The Laplace transform of the time-averaged geometric Brownian
motion in our notation reads \cite{101,102,103}
\begin{eqnarray}
\nonumber
\Upsilon(T;\lambda)&=&\Bigg\langle\exp\left(-\lambda\int^T_0 dt\exp\left(-
\frac{B_t}{a}\right)\right)\Bigg\rangle_{\Psi}\\
\nonumber
&=&\frac{2a}{\sqrt{\pi D_BT}}\int^{\infty}_0dx\exp\left(-\frac{a^2x^2}{D_BT}
\right) \cos\left(2a\sqrt{\frac{\lambda}{D_B}}\sinh(x)\right)\\
&=&\frac{2}{\pi}\int^{\infty}_0dx\exp\left(-\frac{D_BT}{4a^2}x^2\right)
\cosh\left(\frac{\pi x}{2}\right) K_{ix}\left(2a\sqrt{\frac{\lambda}{D_B}}\right),
\label{mgf_geom}
\end{eqnarray}
where $K_{ix}$ is the modified Bessel function of the second kind with purely
imaginary index. As a consequence, the PDF is given by
\begin{eqnarray}
\nonumber
{\bf \Pi}(x,t)&=&\left(\frac{2}{\pi}\right)^{3/2}\int^{\infty}_0dz\exp\left(-
\frac{D_B t}{4 a^2}z^2\right)\cosh\left(\frac{\pi z}{2}\right)\\
\nonumber
&&\times\int^{\infty}_0dw\cos(wx)K_{iz}\left(2a|w|\sqrt{\frac{D_0}{D_B}}\right)\\
&=&\frac{1}{2\sqrt{\pi b_2t(b_1^2+x^2)}}\exp\left(-\frac{{\rm
arcsinh}^2(x/b_1)}{4b_2t}\right),
\label{pdf_3c}
\end{eqnarray}
where
\begin{equation}
b_1=2a\sqrt{\frac{D_0}{D_B}},\quad b_2=\frac{D_B}{4a^2}.
\end{equation}
Recalling that ${\rm arcsinh}(z)\sim z$ for $z\to0$ and ${\rm arcsinh}(z)\sim
\ln(2z)$ for $z\to\infty$, we express the asymptotic behaviour of the PDF as
\begin{equation}
{\bf \Pi}(x,t)\sim\frac{1}{2\sqrt{\pi b_2t(b_1^2+x^2)}}\exp\left(-\frac{x^2}{4b_2t
b_1^2}\right)
\label{87}
\end{equation} 
for $|x|\to0$ and
\begin{equation}
{\bf \Pi}(x,t)\sim\frac{1}{2\sqrt{\pi b_2t(b_1^2+x^2)}}\exp\left(-\frac{\ln^2(2x/b_1)
}{4b_2t}\right)
\label{88}
\end{equation} 
for $|x|\to\infty$.

\begin{figure*}
\centering
\includegraphics[height=\textwidth,angle=270]{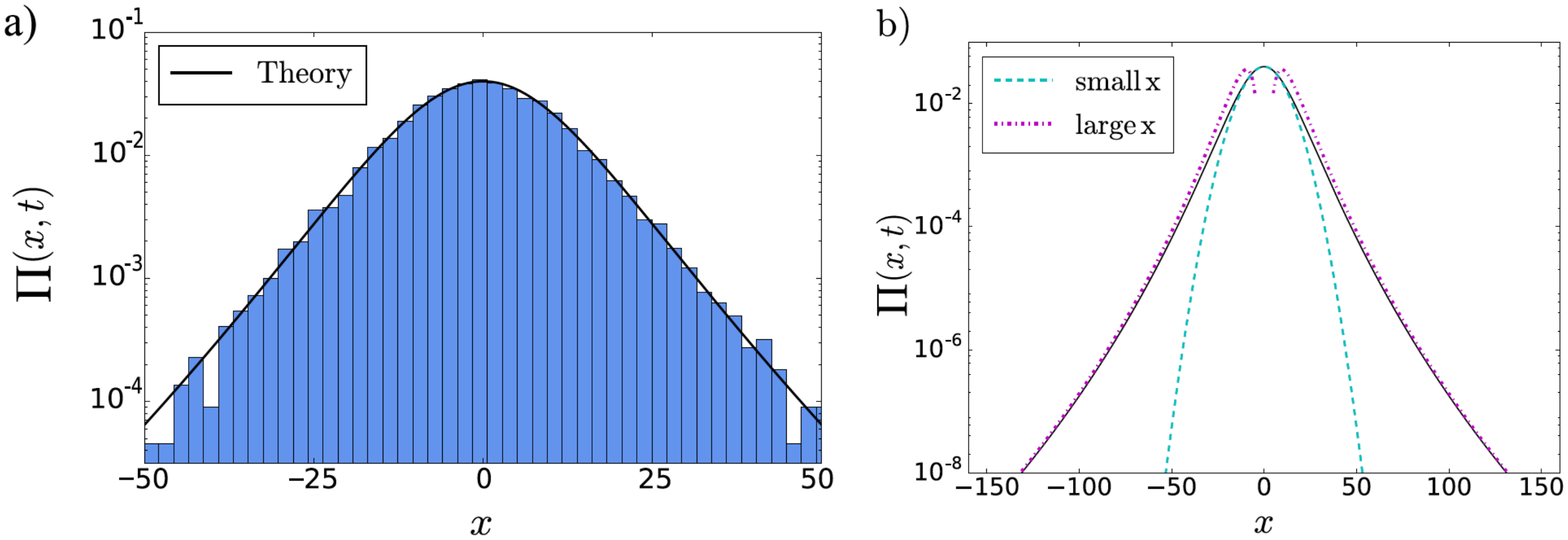}
\caption{PDF at time $t=50$ for diffusivity modelled as geometric Brownian motion,
with $D_B=1$ and $a=10$. Panel a) compares the numerical result and the analytical
expression (\ref{pdf_3c}) (black solid line). Panel b) shows a comparison between
the analytical result and its asymptotic behaviours (\ref{87}) and (\ref{88}).}
\label{fig_prop3}
\end{figure*}

The PDF is shown in Fig. \ref{fig_prop3}. In this case, according to the asymptotic
expansions (\ref{87}) and (\ref{88}) we observe that the central part of the PDF is
approximately Gaussian, while the tails follow a log-normal shape.

\subsection*{Amplitude-PDF $P(A)$}

\begin{figure*}
\centering
\includegraphics[height=\textwidth,angle=270]{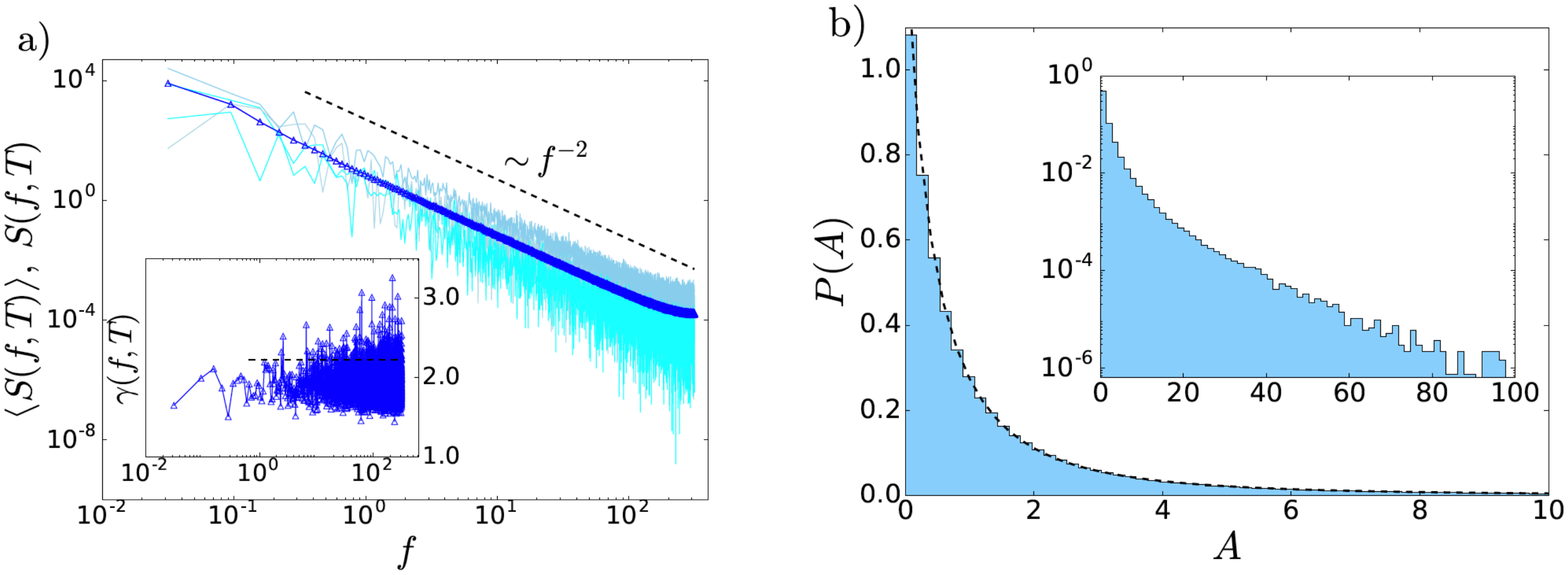}
\caption{Diffusivity modelled as geometric Brownian motion, with $D_B=1$, $a=10$,
and trajectory length $T=10^2$. Panel a) shows the mean power spectrum and a few
realisations of the power spectra from individual trajectories. In the inset the
coefficient of variation is shown---the black dashed line corresponds to the
theoretical result (\ref{gamma_geom}). Panel b) shows the distribution of the
random amplitude $A$. Here the black dashed line corresponds to the numerical evaluation of (\ref{distlargef3}), by making use of (\ref{mgf_geom}) also.}
\label{fig9}
\end{figure*}

Evaluating explicitly $\partial_\lambda\Upsilon(T;\lambda)$
and $\partial^2_\lambda\Upsilon(T;\lambda)$ at $\lambda=0$, we get
\begin{equation}
\gamma=\left[\frac{3}{8}\left(3+e^{D_BT/a^2}\left(2+e^{D_B T/a^2}\right)\right)
-1\right]^{1/2}.
\label{gamma_geom}
\end{equation}
Figure \ref{fig9} summarises our numerical results, which show excellent agreement with the theoretical prediction.

\subsection{Example III: $\Psi_t=B_t^2/a^2$}

Finally, we consider squared Brownian motion
\begin{equation}
\Psi_t=\frac{B_t^2}{a^2},
\end{equation}
where $a$ is a parameter of unit length. Note that here the process $x_t$
in (\ref{lan}) is superdffusive, such that its mean-squared displacement obeys
\begin{equation}
\left \langle \overline{x_t^2} \right \rangle_{\Psi}=\frac{2D_0D_B}{a^2}t^2,
\end{equation} 
and thus a pronounced ageing behaviour occurs.

This example, similarly to the diffusing-diffusivity model in section
\ref{sec3}, defines the diffusivity as the squared of an auxiliary variable,
though in this case the variable follows a Brownian motion instead of the OU
process. This choice implies that there is no crossover time, in contrast to
the standard diffusing-diffusivity model, and thus we obtain a model which
is always non-stationary. In particular, we introduce a larger separation
between small and large values of the diffusivity, which may be interpreted
as non-linear effects of the heterogeneity.  Note that, if we were to define
a random duration $\delta$ of the intervals, this model could be linked to
a correlated CTRW \cite{Tejedor-Metzler,Marcin-Ralf}.

\subsection*{Position-PDF ${\bf \Pi}(x,t)$}

\begin{figure*}
\centering
\includegraphics[height=\textwidth,angle=270]{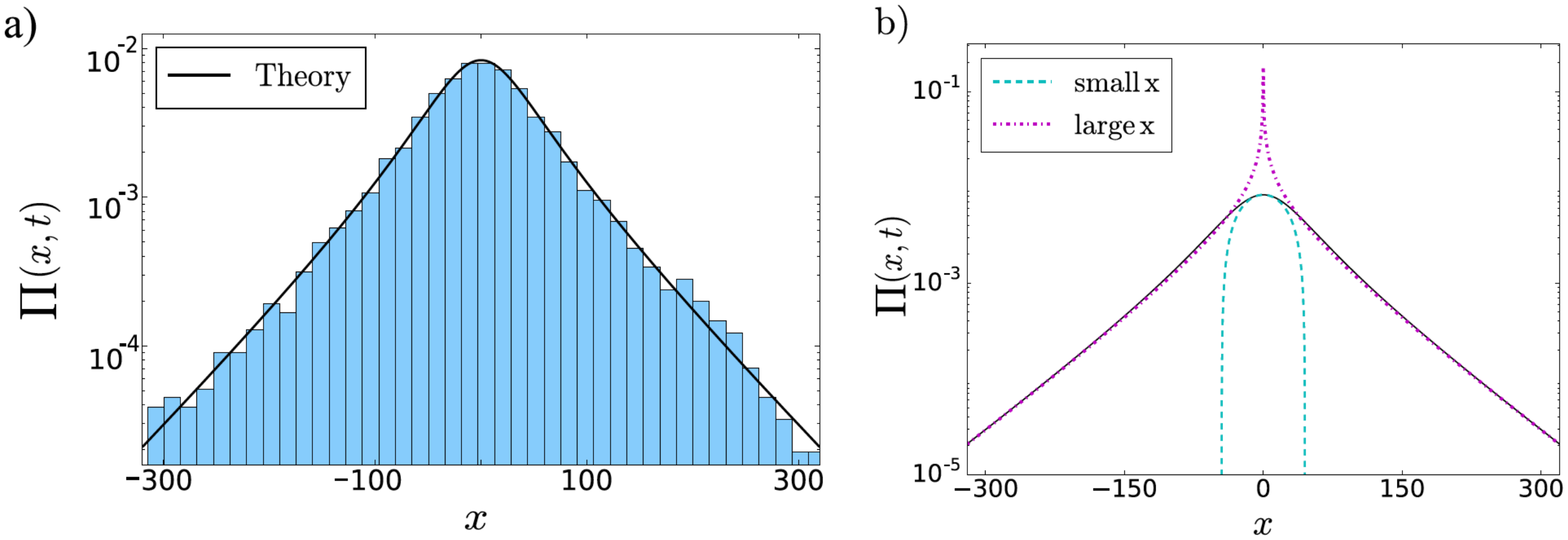}
\caption{Position-PDF at fixed time $t=50$ for diffusivity modelled as squared Brownian motion, with $D_B=1$ and $a=1$. Panel a) compares between the numerical result and expression (\ref{pdf_3a}) (black solid line). Panel b) shows a comparison between the analytical result and its asymptotic behaviour (\ref{63}) and the small $x$ behaviour of ${\rm sech}(z)$ form discussed in the text.}
\label{fig_prop4}
\end{figure*}

The Laplace transform of the PDF of integrated squared Brownian motion was first
calculated in the classical paper by Cameron and Martin \cite{5,5a} (see also
\cite{6}). In our notation,
\begin{eqnarray}
\nonumber
\Upsilon(T;\lambda)&=&\left\langle\exp\left(-\frac{\lambda}{a^2}\int^T_0dtB_t^2
\right)\right\rangle_{\Psi}\\
&=&\frac{\displaystyle 1}{\displaystyle \sqrt{\cosh\left(\sqrt{4D_BT^2\lambda/a^2}\right)}},
\label{tt}
\end{eqnarray}
and the PDF ${\bf \Pi}(x,t)$ for this process is given by
\begin{eqnarray}
\nonumber
{\bf \Pi}(x,t)&=&\frac{1}{\pi}\int^{\infty}_0\frac{dw\cos\left(wx\right)}{\sqrt{
\cosh\left(w\sqrt{4D_BD_0t^2/a^2}\right)}}\\
&=&\frac{1}{\sqrt{2}ct}{\rm sech}\left(\frac{\pi x}{ct}\right)P_{\frac{i
x}{ct}-\frac{1}{2}}\left(0\right),
\label{pdf_3a}
\end{eqnarray}
where $c=2\sqrt{D_0D_B}/a$ and $P_{\nu}(z)$ is the Legendre function of the
first kind. The latter admits the representation
\begin{equation}
P_{ix/ct-1/2}(0)=\sqrt{\pi}\Big/\left[\Gamma\left(\frac{ix}{2ct}+\frac{3}{4}
\right)\Gamma\left(-\frac{ix}{2ct}+\frac{3}{4}\right)\right],
\end{equation}
such that 
\begin{equation}
{\bf \Pi}(x,t)=\frac{\sqrt{\pi}}{ct\sqrt{2}}\frac{{\rm sech}(\frac{\pi x}{ct})}{
\Gamma\left(\frac{ix}{2ct}+\frac{3}{4}\right)\Gamma\left(-\frac{ix}{2ct}+
\frac{3}{4}\right)}.
\end{equation}
In the limits,
\begin{equation}
\Gamma \left(\frac{ix}{2ct}+\frac{3}{4}\right)\Gamma\left(-\frac{ix}{2ct}+
\frac{3}{4}\right)\sim\Gamma\left(\frac{3}{4}\right)^2
\end{equation}
for $|x|\to0$, and
\begin{equation}
\Gamma\left(\frac{ix}{2ct}+\frac{3}{4}\right)\Gamma\left(-\frac{ix}{2ct}+
\frac{3}{4}\right)\sim\pi\sqrt{\frac{2|x|}{ct}}\exp\left(-\frac{\pi|x|}{2ct}\right)
\end{equation}
for $|x|\to\infty$. As a consequence, the behaviour of the PDF for small $x$ is
approximately Gaussian, $\simeq\exp(-\mathrm{const.}x^2)$, where $\mathrm{const.}$
is a constant that can be expressed via the polylogarithm function. Conversely,
\begin{equation}
{\bf \Pi}(x,t)\sim\frac{1}{\pi ct|x|}\exp\left(-\frac{\pi|x|}{2ct}\right)
\label{63}
\end{equation}
for $|x|\to\infty$.
The shape of the PDF is shown in Fig. \ref{fig_prop4}. We clearly observe that the central part is approximately Gaussian while the tails have an exponential trend, as expressed explicitly by the asymptote (\ref{63}).

\begin{figure*}
\centering
\includegraphics[height=\textwidth,angle=270]{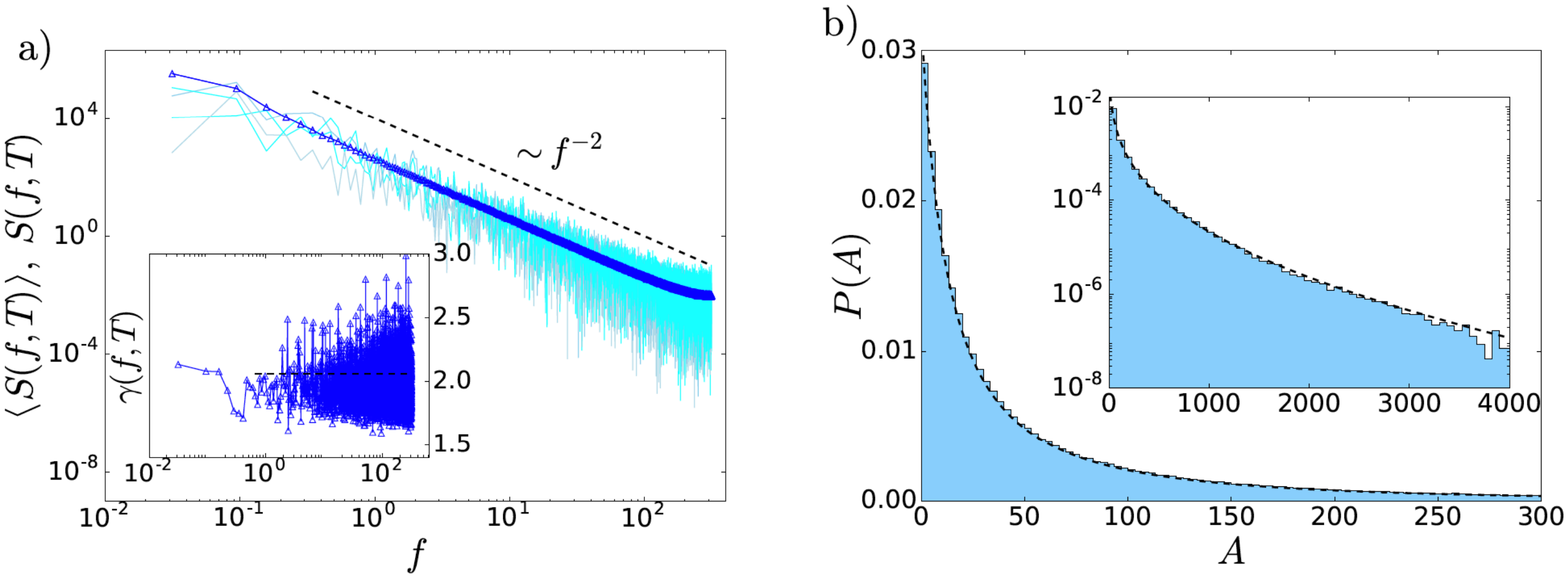}
\caption{Diffusivity modelled as squared Brownian motion, with $D_B=1$, $a=1$,
and trajectory length $T=10^2$. Panel a) shows the mean power spectrum along with
a few realisations of power spectra from individual trajectories. In the inset the coefficient of variation is displayed---the black dashed line indicates the
theoretical value $\sqrt{17}/2\approx2.06$. Panel b) shows the distribution of the random
amplitude $A$. Here the black dashed line corresponds to the numerical evaluation
of (\ref{q}).}
\label{fig7}
\end{figure*}

\subsection*{Amplitude-PDF $P(A)$}

Using (\ref{tt}) we then find that the MGF of the random amplitude $A$ and the
corresponding PDF are given by the integrals
\begin{equation}
\Phi_{\lambda}(A)=\frac{2}{\sqrt{3}}\int^{\infty}_0dp\frac{\displaystyle \exp\left(-4p/3\right)
I_0(2p/3)}{\displaystyle \sqrt{\cosh\left(\sqrt{4D_BT\lambda p/a^2}\right)}}
\end{equation}
and
\begin{equation}
P(A)=\frac{2}{\sqrt{3}}\int^{\infty}_0dz\frac{\displaystyle J_0\left((1+1/\sqrt{3})\sqrt{2Az}
\right)}{\displaystyle \sqrt{\cosh\left(\sqrt{4D_BTz/a^2}\right)}} J_0\left((1-1/\sqrt{3})\sqrt{2Az}\right).
\label{q}
\end{equation}
The series representation of $P(A)$ in (\ref{q}) can be found directly by
taking advantage of expansion (\ref{mu}) in \ref{C} and our
result (\ref{distlargef5}), to yield
\begin{equation}
\fl \qquad P(A)=\frac{1}{2\sqrt{6\pi}} \frac{a^2}{D_BT}\sum_{n=0}^{\infty}\frac{\displaystyle (-1)^n 
\Gamma\left(n+1/2\right)}{\displaystyle \left(n+1/4\right)^2n!(1+ \xi_n)^{3/2}}\,{_2F_1}\left(\frac{3}{4},\frac{5}{4};1;\frac{\xi_n^2}{4(1+\xi_n)^2}\right) 
\label{ZZ}
\end{equation}
with
\begin{equation}
\xi_n=\frac{a^2A}{3\left(n+1/4\right)^2D_BT}.
\end{equation}
Note that the integrals 
\begin{equation}
\int^{\infty}_0d\lambda\lambda^n \Upsilon(T;\lambda)
\end{equation}
exist for any $n>0$ and, hence, all negative moments of $\int^T_0dtB_t^2$ exist,
as well. As a consequence, $P(A)$ is an analytic function of $A$. We immediately
obtain the coefficient of variation from (\ref{gamma}) as $\gamma=\sqrt{17}/2$.
Figure \ref{fig7} displays numerical results for which we observe excellent agreement with the theoretical expressions.

\section{Conclusions}
\label{sec4}

Quite typically stochastic time series are evaluated in terms of the ensemble
averaged MSD $\left \langle \overline{x_t^2}\right \rangle_\Psi$. It has the advantage that fluctuations
are reduced due to the averaging over many individual trajectories. However,
this is not always possible. Namely, for the by-now routine results from
single particle tracking experiments typically rather few, finite time
series are obtained. These are then evaluated in terms of the time-averaged
MSD. While this quantity may also be averaged over the available individual
trajectories, it is increasingly realised that the amplitude fluctuations
between individual trajectories in fact harbours important quantitative
information characteristic for a given stochastic process \cite{pt,pt1,pccp, elipt,greb3}.

Similar to the consideration of time averaged MSDs for trajectories of finite
measurement time $T$, we here analysed the single-trajectory PSD of stochastic
trajectories characterised by random diffusivities. Following our previous
work on standard Brownian motion \cite{we1}, as well as fractional \cite{we2}
and scaled Brownian motion \cite{vittoria}, we here investigated the detailed
behaviour of single-trajectory PSDs of a broad class of diffusing-diffusivity
models. These have recently gained considerable attention as simple models for
diffusion processes in heterogeneous media. We described a general procedure to obtain the position PDF ${\bf \Pi}(x,t)$ for all such models.
The main ingredient in the calculation is the MGF of the integrated diffusivity,
showing explicitly that different choices of the underlying diffusivity $\Psi_t$
lead to distinctly different emerging behaviours, as summarised in Table
\ref{table}).

\begin{table}
\centering
\begin{tabular}{l|l||c|c|c}
 \multicolumn{2}{c||}{  Random diffusivity model} &   MGF of $\tau_T $& ${\bf \gamma}$& Ageing\\
 \specialrule{2pt}{1pt}{1pt}
 \multicolumn{2}{c||}{$\Psi_t=Y_t^2, \quad$ $Y_t=$ OU process } & Eq. (\ref{upsilon1}) &  
 Eq. (\ref{gam_ou}) & $\times$  \\
 \hline
 \multirow{2}{*}{$\Psi_t=\psi_k$} 
 & $\rho(\psi)$ gamma distr. & Eq. (\ref{mgf_gam})   & Eq. (\ref{gam_gam}) & $\times$  \\
 & $\rho(\psi)$ L\'{e}vy-Smirnov distr. & Eq. (\ref{mgf_levy}) & not defined  & not defined   \\
  \hline
 \multirow{3}{*}{$\Psi_t=V[B_t]$} 
  & $V[Z]=\theta(Z)$ & Eq. (\ref{t}) &  $\sqrt{19/8}$ & $\times$   \\
  & $V[Z]=\exp(-Z)$ & Eq. (\ref{mgf_geom}) &  Eq. (\ref{gamma_geom}) & $\checkmark$  \\
  & $V[Z]=Z^2$ & Eq. (\ref{tt}) &  $\sqrt{17}$/2  & $\checkmark$ \\
 \hline
\end{tabular}
\caption{Collection of the main results for the different random diffusivity models
with respective equation numbers. Next to the definitions of the models we refer to
the MGF of the integrated diffusivity $\tau_T$, the coefficient of variation
$\gamma$, which can play the role of an indicator for each model, and the
ageing behaviour.}
\label{table}
\end{table}

We started our discussion from the by-now well-established and widely studied
model of "diffusing-diffusivity", namely the case when $\Psi_t$ is chosen as
the squared of the Ornstein-Uhlenbeck process. We then discussed the second
case, in which $\Psi_t$ is defined as a jump process. The properties of this
model depend strongly on the exact distribution chosen for the increment
variables. We considered two examples, a Gamma distribution and a
L\'{e}vy-Smirnov distribution. Finally, three cases in which the diffusivity
is modelled as a functional of Brownian motion were discussed. 

The main result of this work is that, regardless of the different properties
of all these diffusing-diffusivity models we obtained a universal high-$f$
asymptote of the PSD. This behaviour is characterised by a $1/f^2$ scaling,
in analogy to Brownian motion \cite{we1} and scaled Brownian motion
\cite{vittoria}.
A first way to discriminate among models lies in the study of the ageing
behaviour of the PSD, as already discussed in \cite{vittoria}. Indeed,
we showed that the dependence of the PSD on the trajectory length $T$
appears only for those random diffusivity models that are characterised by
an anomalous scaling of the MSD.
We also showed that differences from one model to another appear in
higher order moments.  In particular, we obtained exact expressions for the
coefficient of variation in all cases, proving that the latter can be a good
indicator of the specific model (see Table \ref{table}).

Finally, we established that the PDF of the random amplitude $A$ carries most
of the meaningful information. Namely, the coefficient of variation may be
directly calculated from its moments. Moreover, its MGF is tightly related to that of the integrated diffusivity, thus reflecting the particular properties of the process $\Psi_t$. As we showed before \cite{we1, we2}, the distribution $P(A)$ can even be evaluated meaningfully from experimental data of fairly short trajectories. In its useful role in data analysis the single-trajectory PSD approach thus complements other methods such as the time-averaged MSD and its amplitude variation \cite{pt,pccp,alek, ralf,samu,samu1,jae_jpa,greb4,greb5}, ageing analyses \cite{we2,pccp}, or, in the context of non-Gaussian diffusion, the codifference methods \cite{jakub}.

The role of distinguishing different physical processes from measured single
trajectory data therefore heavily lies on the amplitude fluctuations and the
coefficient of variation encoded in them. This improved understanding of the
single-trajectory PSD should therefore replace a common claim in many textbooks
according to which the $1/f^2$-dependence of the spectrum was a fingerprint of
Brownian motion. In line to previous works \cite{we1,we2,vittoria}, in which we
already alerted that this may be a deceptive concept, we have shown here that a
wide range of random-diffusivity models with distinctly different behaviour and
showing anomalous diffusion, exhibit precisely the $1/f^2$-dependence. Therefore,
any experimental observation of the spectral density varying as $1/f^2$ alone
cannot be taken as proof of standard Brownian motion. One necessarily needs to
consider the ageing behaviour of the spectral density, as in the case of
superdiffusive fractional Brownian motion or scaled Brownian motion, evaluate
the coefficient of variation of the spectral density, or determine the functional
form of the PDF $P(A)$ of the amplitude fluctuations.

In light of this the relevance of the presented results is twofold. First,
they provide new and useful insights into the increasingly popular class
of stochastic processes with random diffusivity used in the description
of Fickian yet non-Gaussian diffusion in heterogeneous systems. Second,
the results continue our ongoing analysis based on the single-trajectory
PSD for different classes of stochastic processes, showing in particular
the persistence of the $1/f^2$-scaling of the PSD, which appears to be
robust---as long as we do not introduce correlations in the driving noise
of the system, as studied in \cite{we2}.

We note that if we redefine $\dot{x}_t$ in the Langevin equation (\ref{lan})
as $\dot{S}_t/S_t$, we recover the seminal Black-Scholes (or Black-Scholes-Merton)
equation for an asset price $S_t$ with zero-constant trend and stochastic volatility
$\sqrt{D_0\Psi_t}$ used in financial market models \cite{blac73,mert71,mert75}. The
relevance of diffusing-diffusivity approaches to economic and financial modelling was
also discussed
for the case of the squared Ornstein-Uhlenbeck process \cite{Chechkin:DD2}. Namely,
the resulting stochastic equation for $D_0\Psi_t$ in this case is nothing else than
the Heston model \cite{heston}, a special class of the Cox-Ingersson-Ross model
\cite{cir,cir1}, and as such specifies the time evolution of the stochastic volatility
of a given asset \cite{Lanoiselee18a,heston,yako}.

We finally note that realisation-to-realisation fluctuations of a stochastic
process also turn out to be relevant in many scenarios of first-passage time
statistics \cite{carlos,carlos1}. These fluctuations are connected to the
typically broad ("defocused") PDFs of first-passage times even in simple
geometries \cite{aljaz,denis}. It will be interesting to extend the existing
first-passage time analyses of diffusing-diffusivity models \cite{Lanoiselee18b, Sposini19} to the different processes studied herein, and to more complex
geometries.

\section*{Acknowledgments}
VS is supported by the Basque Government through the BERC 2018-2021 program 
and by the Spanish Ministry of Economy and Competitiveness MINECO through BCAM 
Severo Ochoa accreditation SEV-2017-0718.
FS acknowledges Padova University for support within grant PRD-BIRD191017.
RM acknowledges Deutsche Forschungsgemeinschaft (DFG) for support within
grant ME 1535/7-1, as well as the Foundation for Polish Science (Fundacja
na rzecz Nauki Polskiej) for an Alexander von Humboldt Polish Honorary
Research Scholarship.

\appendix

\section{Derivation of the moment-generating function of the power spectral
density}
\label{app}

It is convenient to rewrite formally the definition of the PSD in (\ref{1}) in
the form
\begin{equation}
\label{1ap}
S_T(f)=\frac{1}{T}\left(\int^T_0dt\cos(ft)x_t\right)^2+\frac{1}{T}\left(\int^T_0
dt\sin(ft)x_t\right)^2. 
\end{equation}
Our first step then consists in a standard linearisation of the expression in the
exponential in (\ref{b}). Taking advantage of the integral identity
\begin{equation}
\exp\bigl(-b^2/(4c)\bigr)=\sqrt{\frac{c}{\pi}}\int^{\infty}_{-\infty}dz\exp(-c
z^2+ibz)
\end{equation}
for $c>0$, we formally recast (\ref{b}) into the form
\begin{equation}
\fl \qquad
\phi_{\lambda}=\frac{1}{4\pi\lambda}\int^{\infty}_{-\infty}dz_1\int^{\infty}
_{-\infty}dz_2\exp\left(-\frac{z_1^2+z_2^2}{4\lambda}\right)
\left\langle\overline{\exp\left(i\int^T_0dtQ_tx_t\right)}\right\rangle_{\Psi},
\label{3}
\end{equation}
where $Q_t$ is defined in (\ref{Q}). Now, the averaging over thermal noise
realisations can be performed straightforwardly, yielding
\begin{eqnarray}
\nonumber
\overline{\exp\left(i\int^T_0dtQ_tx_t\right)}
&=&\overline{\exp\left(i\sqrt{2D_0}\int^T_0dt\sqrt{\Psi_t}\xi_t\int^T_t d\tau
Q_{\tau}\right)}\\
&=&\exp\left(-D_0\int^T_0dt\Psi_t\left(\int^T_t d\tau Q_{\tau}\right)^2\right),
\label{zz}
\end{eqnarray}
where we integrated by parts and used (\ref{lan}). Combining (\ref{zz}) and
(\ref{3}) we arrive at our general result (\ref{4}).

The derivation of our main result (\ref{main}) takes advantage of the explicit
form of $Q$ in (\ref{Q}) and the following calculation,

\begin{eqnarray}
\nonumber
\fl  \qquad \int^T_0dt\Psi_t & \left(\int^T_t d\tau Q_{\tau}\right)^2 = 
\frac{z_1^2}{f^2T} \int^T_0dt\Psi_t\bigg(\sin(fT)-\sin(ft)\bigg)^2 \\
\nonumber
\fl & +\frac{z_2^2}{f^2T}\int^T_0dt \Psi_t\bigg(\cos(ft)-\cos(fT)\bigg)^2\\
\fl &+\frac{2z_1z_2}{f^2T} \int^T_0 dt \Psi_t \bigg( \sin(fT)-\sin(ft) \bigg) \bigg(\cos(ft)-\cos(fT)\bigg).
\end{eqnarray}
Inserting the latter expression into (\ref{4}) and performing the integrations
over $z_1$ and $z_2$ we find the expression in (\ref{main}) with $L_f(t_1,t_2)$
explicitly defined by

\begin{eqnarray}
\label{L}
\nonumber
\fl L_f & (t_1,t_2)=\frac{1}{2}\cos(2ft_2)-\frac{1}{2}\cos(2ft_1)-\frac{1}{2}\cos(f(T-t_1))-\frac{1}{2}\cos(f(T-t_2)) \\
\nonumber
\fl & -\frac{1}{4}\cos(2f(T-t_1)) -\frac{1}{2}\cos(2f(T-t_2)) +\frac{1}{4}\cos(f(3T-t_2))\\
\nonumber 
\fl &-\frac{1}{4}\cos(f(3T-t_1)) +\frac{3}{4}\cos(f(T+t_1)) -\frac{3}{4}\cos(f(T+t_2))-\frac{1}{2}\cos(f(t_1-t_2))\\
\nonumber 
\fl &-\frac{1}{4}\cos(2f(t_1-t_2)) +\frac{1}{4}\cos(f(T-2t_1-t_2)) -\frac{1}{4} \cos(f(T-t_1-2t_2))\\
\nonumber
\fl & +\frac{1}{2}\cos(f(T+t_1-2 t_2)) +\frac{1}{2}\cos(f(T-2t_1+t_2))+\frac{1}{4}\cos(f(T+2t_1-t_2))\\
\nonumber
\fl &-\frac{1}{4} \cos(f(T-t_1+2t_2)) +\frac{1}{2}\cos(f(2T-t_1-t_2))-\frac{1}{2}\cos(f(2T+t_1-t_2)) \\
\fl & +\frac{1}{2}\cos(f(2T-t_1+t_2)).
\end{eqnarray}

\section{High-$f$ behaviour of the Riemann-integrable diffusivities}
\label{B}

For Riemann-integrable functions, according to the Riemann-Lebesgue lemma we
have
\begin{equation}
\label{c1}
\lim_{f\to\infty}\int^T_0dt\Psi_t\cos(f(T-t))=0
\end{equation}
with probability $1$ (however, nothing can be said about how fast zero is
approached in the general case). Once (\ref{c1}) holds, one finds that for
$L_f(t_1,t_2)$ defined in (\ref{L}), 
\begin{equation}
\lim_{f\to\infty}\int^T_0dt_1\Psi_{t_1}\int^T_0dt_2\Psi_{t_2}L_f(t_1,t_2)=0.
\end{equation}

\section{Useful formulae}
\label{C}

Our expression for the PDF $P(A)$ in (\ref{Bess}) and (\ref{zero}) rely on
the following series expansion of the product of two Bessel functions,
\begin{eqnarray}
\nonumber
&J_0&\left(\left(1+\frac{1}{\sqrt{3}}\right)\sqrt{2zA}\right)J_0\left(\left(1
-\frac{1}{\sqrt{3}}\right)\sqrt{2zA}\right)\\
&&=\sum_{n=0}^{\infty}\frac{(-1)^n}{(n!)^2}\left(\frac{\sqrt{3}+1}{\sqrt{6}}
\right)^{2n} \,_2F_1\left(-n,-n;1;\frac{1-\sqrt{3}/2}{1+\sqrt{3}/2}\right)(zA)^n.
\label{Bes}
\end{eqnarray}
The form of the PDF in (\ref{ZZ}) stems from the expansion
\begin{equation}
\fl \qquad \frac{\displaystyle 1}{\displaystyle \sqrt{\cosh\left(\sqrt{4D_BTz/a^2}\right)}}=\sqrt{2}
\sum_{n=0}^{\infty} {{-1/2}\choose{n}}\exp\left(-2\sqrt{\frac{D_BTz}{a^2}}\left(2n+\frac{1}{2}
\right)\right).
\label{mu}
\end{equation}
The result in (\ref{smallS}) involves toroidal functions defined by
\begin{equation}
\fl \qquad
P_{n-1/2}\left(\cosh(\eta)\right)=\frac{\displaystyle 2}{\displaystyle \pi}e^{-(n+1/2)\eta}
\int^{\pi/2}_0\frac{\displaystyle d\phi}{\displaystyle \left(1-2e^{-\eta}\sinh(
\eta)\sin^2(\phi)\right)^{n+1/2}}.
\label{tor}
\end{equation}
Setting $\exp(-\eta)\sinh(\eta)=1/3$, i.e., $\eta=\ln(\sqrt{3})$, we obtain
expression (\ref{smallS}).

\section*{References}

\bibliographystyle{iopart-num}

\begin{thebibliography}{99}

\bibitem{brown} R. Brown, \emph{A brief account of microscopical observations made in the months of June, July and August 1827, on the particles contained in the pollen of plants; and on the general existence of active molecules in organic and inorganic bodies}, Phil. Mag. \textbf{4}, 161 (1828).

\bibitem{fick} A. Fick,
\emph{{\"U}ber Diffusion (On Diffusion)},
Ann. Phys. (Leipzig) \textbf{170}, 59 (1855).

\bibitem{einstein} A. Einstein,
\emph{{\"U}ber die von der molekularkinetischen Theorie der W{\"a}rme
geforderte Bewegung von in ruhenden Fl{\"u}ssigkeiten suspendierten Teilchen
(On the motion of small particles suspended in liquids at rest required by
the molecular-kinetic theory of heat)},
Ann. Phys. (Leipzig) {\bf 322}, 549 (1905).

\bibitem{smoluchowski} M. von Smoluchowski,
\emph{Zur kinetischen Theorie der Brownschen Molekularbewegung und der
Suspensionen (On the kinetic theory of Brownian molecular motion and suspensions)},
Ann. Phys. (Leipzig) \textbf{21}, 756 (1906).

\bibitem{sutherland} W. Sutherland,
\emph{A dynamical theory of diffusion for non-electrolytes and the molecular
mass of albumin},
Philos. Mag. \textbf{9}, 781 (1905).

\bibitem{pearson} K. Pearson,
\emph{The Problem of the Random Walk},
Nature \textbf{72}, 294 (1905).

\bibitem{langevin} P. Langevin,
\emph{On the theory of Brownian motion},
C. R. Acad. Sci. Paris \textbf{146}, 530 (1908).

\bibitem{hoefling} F. H\"ofling and T. Franosch,
\emph{Anomalous transport in the crowded world of biological cells},
Rep. Progr. Phys. \textbf{76}, 046602 (2013).

\bibitem{lene} K. N{\o}rregaard, R. Metzler, C. M. Ritter, K. Berg-S{\o}rensen,
and L. B. Oddershede,
\emph{Manipulation and motion of organelles and single molecules in living cells},
Chem. Rev. \textbf{117}, 4342 (2017).

\bibitem{xie} X. S. Xie, P. J. Choi, G.-W. Li, N. K. Lee, and G. Lia,
\emph{Single-Molecule Approach to Molecular Biology in Living Bacterial Cells},
Annu. Rev. Biophys. \textbf{37}, 417 (2008).

\bibitem{brauchle} C. Br{\"a}uchle, D. C. Lamb, and J. Michaelis, Single Particle
Tracking and Single Molecule Energy Transfer (Wiley-VCH, Weinheim, Germany,
2012).

\bibitem{ilpo} M. Javanainen H. Martinez-Seara, R. Metzler, and I. Vattulainen,
\emph{Diffusion of Integral Membrane Proteins in Protein-Rich Membranes},
J. Phys. Chem. Lett. \textbf{8}, 4308 (2017).

\bibitem{smith} X. Hu, L. Hong, M. D. Smith, T. Neusius, X. Cheng, and J. C. Smith,
\emph{The dynamics of single protein molecules is non-equilibrium and self-similar
over thirteen decades in time},
Nature Phys. \textbf{12}, 171 (2016).

\bibitem{pt} E. Barkai, Y. Garini, and R. Metzler,
\emph{Strange Kinetics of Single Molecules in Living Cells},
Phys. Today \textbf{65(8)}, 29 (2012).

\bibitem{pt1} D. Krapf and R. Metzler,
\emph{Strange interfacial molecular dynamics},
Phys. Today \textbf{72(9)}, 48 (2019).

\bibitem{statphys} R. Metzler,
\emph{Brownian motion and beyond: first-passage, power spectrum,
non-Gaussianity, and anomalous diffusion},
J. Stat. Mech. \textbf{(2019)} 114003.

\bibitem{yossipt} J. Klafter, M. F. Shlesinger, and G. Zumofen,
\emph{Beyond Brownian Motion},
Phys. Today \textbf{49(6)}, 33 (1996).

\bibitem{mikenature} M. Shlesinger, G. Zaslavsky, and J. Klafter,
\emph{Strange kinetics},
Nature \textbf{363}, 31 (1993).

\bibitem{harveypt} H. Scher, M. F. Shlesinger, and J. T. Bendler,
\emph{Time-Scale Invariance in Transport and Relaxation},
Phys. Today \textbf{44(1)}, 26 (1991).

\bibitem{saxton} M. J. Saxton and K. Jacobsen,
\emph{Single-particle tracking: applications to membrane dynamics},
Ann. Rev. Biophys. Biomol. Struct. \textbf{26}, 373 (1997).

\bibitem{golding} I. Golding and E. C. Cox,
\emph{Physical Nature of Bacterial Cytoplasm},
Phys. Rev. Lett. \textbf{96}, 098102 (2006).

\bibitem{weber} S. C. Weber, A. J. Spakowitz, and J. A. Theriot,
\emph{Bacterial Chromosomal Loci Move Subdiffusively through a Viscoelastic Cytoplasm},
Phys. Rev. Lett. \textbf{104}, 238102 (2010).

\bibitem{yuval} K. Burnecki, E. Kepten, J. Janczura, I. Bronshtein, Y. Garini,
and A. Weron,
\emph{Universal Algorithm for Identification of Fractional Brownian Motion. A
Case of Telomere Subdiffusion},
Biophys. J. \textbf{103}, 1839 (2012).

\bibitem{yuval1} I. Bronstein, Y. Israel, and Y. Garini,
\emph{Transient anomalous diffusion of telomeres in the nucleus of mammalian cells},
Phys. Rev. Lett. \textbf{103}, 018102 (2009).

\bibitem{leneprl} J.-H. Jeon, V. Tejedor, S. Burov, E. Barkai, C. Selhuber-Unkel,
K. Berg-S{\o}rensen, L. Oddershede, and R. Metzler,
\emph{In vivo anomalous diffusion and weak ergodicity breaking of lipid granules},
Phys. Rev. Lett. \textbf{106}, 048103 (2011).

\bibitem{greb} E. Bertseva, D. S. Grebenkov, P. Schmidhauser, S. Gribkova,
S. Jeney, and L. Forro,
\emph{Optical Trapping Microrheology in Cultured Human Cells},
Eur. Phys. J. E \textbf{35}, 63 (2012).

\bibitem{diego} A. V. Weigel, B. Simon, M. M. Tamkun, and D. Krapf,
\emph{Ergodic and nonergodic processes coexist in the plasma membrane as observed
by single-molecule tracking},
Proc. Natl. Acad. Sci. USA  \textbf{108}, 6438 (2011).

\bibitem{carlo} C. Manzo, J. A. Torreno-Pina, P. Massignan, G. J. Lapeyre, Jr.,
M. Lewenstein, and M. F. Garcia Parajo,
\emph{Weak ergodicity breaking of receptor motion in living cells stemming from
random diffusivity},
Phys. Rev. X \textbf{5}, 011021 (2015).

\bibitem{matthias} J. Szymanski and M. Weiss,
\emph{Elucidating the origin of anomalous diffusion in crowded fluids},
Phys. Rev. Lett. \textbf{103}, 038102 (2009).

\bibitem{lene1} J.-H. Jeon, N. Leijnse, L. B. Oddershede, and R. Metzler,
\emph{Anomalous diffusion and power-law relaxation in wormlike micellar solution},
New J. Phys. \textbf{15}, 045011 (2013).

\bibitem{gratton} C. Di Rienzo, V. Piazza, E. Gratton, F. Beltram, and F. Cardarelli,
\emph{Probing short-range protein Brownian motion in the cytoplasm of living cells},
Nature Comm. \textbf{5}, 5891 (2014).

\bibitem{granick} K. Chen, B. Wang, and S. Granick,
\emph{Memoryless self-reinforcing directionality in endosomal active transport
within living cells},
Nature Mat. \textbf{14}, 589 (2015).

\bibitem{roberts} D. Robert, T. H. Nguyen, F. Gallet, and C. Wilhelm,
\emph{In vivo determination of fluctuating forces during endosome trafficking
using a combination of active and passive microrheology},
PLoS ONE \textbf{4}, e10046 (2010).

\bibitem{elbaum} A. Caspi, R. Granek, and M. Elbaum,
\emph{Enhanced diffusion in active intracellular transport},
Phys. Rev. Lett. \textbf{85}, 5655 (2000).

\bibitem{jaenature} M. S. Song, H. C. Moon, J.-H. Jeon, and H. Y. Park,
\emph{Neuronal messenger ribonucleoprotein transport follows an aging L{\'e}vy walk},
Nat. Comm. \textbf{9}, 344 (2018).

\bibitem{christine} J. F. Reverey, J.-H. Jeon, H. Bao, M. Leippe, R. Metzler, and C.
Selhuber-Unkel,
\emph{Superdiffusion dominates intracellular particle motion in the supercrowded
space of pathogenic Acanthamoeba castellanii},
Sci. Rep. \textbf{5}, 11690 (2015).

\bibitem{granicknm} B. Wang, J. Kuo, S. C. Bae, and S. Granick,
\emph{When Brownian diffusion is not Gaussian},
Nat. Mater. \textbf{11}, 481 (2012).

\bibitem{granick1} B. Wang, S. M. Anthony, S. C. Bae, and S. Granick,
\emph{Anomalous yet Brownian},
Proc. Natl. Acad. Sci. USA \textbf{106}, 15160 (2009).

\bibitem{granick2} J. Guan, B. Wang, and S. Granick,
\emph{Single-molecule observation of long jumps in polymer adsorption},
ACS Nano \textbf{8}, 3331 (2014).

\bibitem{post} K. He, F. B. Khorasani, S. T. Retterer, D. K. Tjomasn, J. C.
Conrad, and R. Krishnamoorti,
\emph{Diffusive dynamics of nanoparticles in arrays of nanoposts},
ACS Nano \textbf{7}, 5122 (2013).

\bibitem{xue:BYNG8} C. Xue, X. Zheng, K. Chen, Y. Tian, and G. Hu,
\emph{Probing non-Gaussianity in confined diffusion of nanoparticles},
J. Phys. Chem. Lett. {\bf 7}, 514 (2016).

\bibitem{xue1} D. Wang, R. Hu, M. J. Skaug, and D. Schwartz,
\emph{Temporally anticorrelated motion of nanoparticles at a liquid interface},
J. Phys. Chem. Lett. {\bf 6} 54 (2015).

\bibitem{xue2} S. Dutta and J. Chakrabarti,
\emph{Anomalous dynamical responses in a driven system},
EPL  {\bf 116}, 38001 (2016).

\bibitem{Goldstein:BYNG9} K. C. Leptos, J. S. Guasto, J. P. Gollub, A. I.
Pesci, and R. E. Goldstein,
\emph{Dynamics of enhanced tracer diffusion in suspensions of swimming
eukaryotic microorganisms},
Phys. Rev. Lett. {\bf 103}, 198103 (2009).

\bibitem{hapca} S. Hapca, J. W. Crawford, and I. M. Young,
\emph{Anomalous diffusion of heterogeneous populations characterized by normal
diffusion at the individual level},
J. R. Soc. Interface {\bf 6}, 111 (2009).

\bibitem{greb1} P. Witzel, M. G{\"o}tz, Y. Lanoisel{\'e}e, T. Franosch, D. S.
Grebenkov, and D. Heinrich, \emph{Heterogeneities Shape Passive Intracellular
Transport}, Biophys. J. \textbf{117}, 203 (2019).

\bibitem{beta} A. G. Cherstvy, O. Nagel, C. Beta, and R. Metzler,
\emph{Non-Gaussianity, population heterogeneity, and transient superdiffusion
in the spreading dynamics of amoeboid cells}, Phys. Chem. Chem. Phys.,
\textbf{20}, 23034 (2018).

\bibitem{ilpo1} J.-H. Jeon, M. Javanainen, H. Martinez-Seara, R. Metzler, and I.
Vattulainen, \emph{Protein crowding in lipid bilayers gives rise to non-Gaussian
anomalous lateral diffusion of phospholipids and proteins}, Phys. Rev. X
\textbf{6}, 021006 (2016).

\bibitem{beck} C. Beck and E. G. D. Cohen,
\emph{Superstatistics},
Physica A \textbf{332}, 267 (2003).

\bibitem{beck1} C. Beck,
\emph{Statistics of three-dimensional Lagrangian turbulence},
Phys. Rev. Lett. \textbf{98}, 064502 (2007).

\bibitem{beck2} C. Beck,
\emph{Stretched exponentials from superstatistics},
Physica A \textbf{365}, 96 (2006).

\bibitem{beck3} E. van der Straeten and C. Beck,
\emph{Superstatistical fluctuations in time series: Applications to share-price
dynamics and turbulence},
Phys. Rev. E \textbf{80}, 036108 (2009).

\bibitem{erice} R. Metzler,
\emph{Superstatistics and non-Gaussian diffusion},
Euro. Phys. J. Special Topics (at press).

\bibitem{fulvio} F. Baldovin, E. Orlandini, and F. Seno, \emph{Polymerization
induces non-Gaussian diffusion}, Frontiers Phys. \textbf{7}, 124 (2019) .

\bibitem{gianni} A. Mura, M. S. Taqqu, and F. Mainardi,
\emph{Non-Markovian diffusion equations and processes: analysis and simulations},
Physica A {\bf 387}, 5033 (2008).

\bibitem{gianni1} A. Mura and G. Pagnini,
\emph{Characterizations and simulations of a class of stochastic processes to
model anomalous diffusion},
J. Phys. A {\bf 41}, 285003 (2008).

\bibitem{gianni2} D. Molina-Garc{\'i}a, T. Minh Pham, P. Paradisi, C. Manzo,
and G. Pagnini,
\emph{Fractional kinetics emerging from ergodicity breaking in random media},
Phys. Rev. E \textbf{94}, 052147 (2016).

\bibitem{Chubynsky14} Chubynsky MV and Slater GW,
{\em Diffusing Diffusivity: A Model for Anomalous, yet Brownian, Diffusion},
Phys. Rev. Lett. {\bf 113}, 098302 (2014).

\bibitem{Jain16} R. Jain and K. L. Sebastian, 
{\em Diffusion in a Crowded, Rearranging Environment},
J. Phys. Chem. B {\bf 120}, 3988 (2016).

\bibitem{Jain16b} R. Jain and K. L. Sebastian,
{\em Diffusing diffusivity: survival in a crowded rearranging and bounded domain},
J. Phys. Chem. B {\bf 120}, 9215 (2016).

\bibitem{Chechkin:DD2} A. V. Chechkin, F. Seno, R. Metzler, and I. M. Sokolov,
{\em Brownian Yet Non-Gaussian Diffusion: From Superstatistics to Subordination
of Diffusing Diffusivities},
Phys. Rev. X {\bf 7}, 021002 (2017).

\bibitem{cherayil} N. Tyagi and B. J. Cherayil,
\emph{Non-Gaussian Brownian diffusion in dynamically disordered thermal
environments},
J. Phys. Chem. B {\bf 121}, 7204 (2017).

\bibitem{Lanoiselee18a}	Y. Lanoisel{\'e}e and D. S. Grebenkov,
{\em A model of non-Gaussian diffusion in heterogeneous media},
J. Phys. A. {\bf 51}, 145602 (2018).

\bibitem{Lanoiselee18b}	Y. Lanoisel{\'e}e, N. Moutal, and D. S. Grebenkov,
{\em Diffusion-limited reactions in dynamic heterogeneous media},
Nature Commun. {\bf 9}, 4398 (2018).

\bibitem{Sposini18} V. Sposini, A. V. Chechkin, F. Seno, G. Pagnini, and R. Metzler,
{\em Random diffusivity from stochastic equations: comparison of two models for
Brownian yet non-Gaussian diffusion,}
New. J. Phys. {\bf 20}, 043044 (2018).

\bibitem{Grebenkov19} D. S. Grebenkov, 
{\em A unifying approach to first-passage time distributions in diffusing
diffusivity and switching diffusion models,} 
J. Phys. A {\bf 52}, 174001 (2019). 

\bibitem{Lanoiselee19} Y. Lanoisel{\'e}e and D. S. Grebenkov, 
{\em Non-Gaussian diffusion of mixed origins,} 
J. Phys. A {\bf 52}, 304001 (2019). 

\bibitem{Sposini19} V. Sposini, A. V. Chechkin, and R. Metzler,
{\em First passage statistics for diffusing diffusivity,}
J. Phys. A {\bf 52}, 04LT01 (2019).
			
\bibitem{Barkai19} M. Hidalgo-Soria and E. Barkai, 
{\em The Hitchhiker model for Laplace diffusion processes in the cell environment,}
E-print arXiv:1909.07189.
			
\bibitem{Roichman19} I. Chakraborty and Y. Roichman, 
{\em Two coupled mechanisms produce Fickian, yet non-Gaussian diffusion in
heterogeneous media} 
E-print arXiv:1909.11364 (2019).
		
\bibitem{Shephard} N. Shephard, {\em Stochastic volatility models.} In: S. N. Durlauf,
L.E. Blume (eds) Macroeconometrics and Time Series Analysis. The New Palgrave Economics 
Collection. Palgrave Macmillan, London (2010).		

\bibitem{Barndorff-Nielsen} O. Barndorff-Nielsen and N. Shephard, {\em Non-Gaussian OU
based models and some of their uses in financial economics}, J. R. Stat. Soc. B {\bf 63},
167--241 (2001).
		 			
\bibitem{Krapf14} S. Sadegh, E. Barkai and D. Krapf,
{\em 1/f noise for intermittent quantum dots exhibits non-stationarity and
critical exponents},
New J. Phys. {\bf 16}, 113054 (2014).
			
\bibitem{Barkai16} N. Leibovich,A. Dechant, E. Lutz, and E. Barkai,
{\em Aging Wiener-Khinchin Theorem and Critical Exponents of $1/f$ Noise},
Phys. Rev. E {\bf 94}, 052130 (2016).
			
\bibitem{Barkai17} N. Leibovich and E. Barkai,
{\em Conditional $1/f^\alpha$ noise: From single molecules to macroscopic
measurement},
Phys. Rev. E {\bf 96}, 032132 (2017).

\bibitem{we1} D. Krapf, E. Marinari, R. Metzler, G. Oshanin, X. Xu and A. Squarcini, 
{\em Power spectral density of a single Brownian trajectory: what one can and
cannot learn from it},  
New J. Phys. {\bf 20}, 023029 (2018).
			
\bibitem{we2} D. Krapf, N. Lukat, E. Marinari, R. Metzler, G. Oshanin, C.
Selhuber-Unkel, A. Squarcini, L. Stadler, M. Weiss, and X. Xu, 
{\em Spectral Content of a Single Non-Brownian Trajectory}, 
Phys. Rev. X {\bf 9}, 011019 (2019). 
			
\bibitem{vittoria} V. Sposini, R. Metzler and G. Oshanin, 
{\em Single-trajectory spectral analysis of scaled Brownian motion,}
New J. Phys. {\bf 21}, 073043 (2019).

\bibitem{sbm} S. C. Lim and S. V. Muniandy,
\emph{Self-similar Gaussian processes for modeling anomalous diffusion},
Phys. Rev. E \textbf{66}, 021114 (2002).

\bibitem{sbm1} J.-H. Jeon, A. V. Chechkin, and R. Metzler,
\emph{Scaled Brownian motion: a paradoxical process with a time dependent
diffusivity for the description of anomalous diffusion},
Phys. Chem. Chem. Phys. \textbf{16}, 15811 (2014).

\bibitem{jakub1} J. \'Sl\c{e}zak, K. Burnecki, and R. Metzler,
\emph{Random coefficient autoregressive processes describe Brownian yet
non-Gaussian diffusion in heterogeneous systems},
New J. Phys. \textbf{21}, 073056 (2019).

\bibitem{Feller51} W. Feller,
{\em Two singular diffusion problems}, 
Ann. Math. {\bf 54}, 173 (1951).

\bibitem{Cox85} J. C. Cox, J. E. Ingersoll, and S. A. Ross,
{\em A theory of the term structure of interest rates},
Econometrica {\bf 53}, 385 (1985).

\bibitem{Heston93} S.L. Heston, {\em A Closed-Form Solution for Options with
Stochastic Volatility with Applications to Bond and Currency Options}, 
Rev. Financial Studies {\bf 6}, 327 (1993).

\bibitem{Dankel} T. Dankel, 
{\em On the distribution of the integrated square of the Ornstein-Uhlenbeck
process},
SIAM J. Appl. Math. {\bf 5}, 568 (1991).

\bibitem{Tejedor-Metzler} V. Tejedor and R. Metzler,
{\em Anomalous diffusion in correlated continuous time random walks},
J. Phys. A {\bf 43}, 082002 (2010).

\bibitem{Marcin-Ralf} M. Magdziarz, R. Metzler, W. Szczotka, and P. Zebrowski,
{\em Correlated continuous-time random walks -- scaling limits and Langevin
picture}, J. Stat. Mech. P04010 (2012).

\bibitem{0} P. L\'evy, 
{\em Sur certains processus stochastiques homog\`{e}nes (On certain homogeneous
stochastic processes)}, 
Compositio Mathematica {\bf 7}, 283 (1940).

\bibitem{5} R. H. Cameron and W. T. Martin, 
{\em Transformations of Wiener integrals under a general class of linear
transformation},
Trans. Amer. Math. Soc. {\bf 58}, 184 (1945);

\bibitem{5a} R. H. Cameron and W. T. Martin,
{\em Evaluation of various Wiener integrals by use of certain Sturm-Liouville
differential equations}, 
Bull. Amer. Math. Soc. {\bf 51}, 73 (1945).

\bibitem{6} M. Kac, 
{\em On distributions of certain Wiener functionals}, 
Trans. Amer. Math. Soc. {\bf 65}, 1 (1949).

\bibitem{7} P. Erd\"os and M. Kac, 
{\em On the number of positive sums of independent random variables},
Bull. Amer. Math. Soc. {\bf 53}, 1011 (1947).  

\bibitem{8} J. Lamperti, 
{\em An occupation time theorem for a class of stochastic processes}, 
Trans. Amer. Math. Soc. {\bf 88}, 380 (1958).

\bibitem{1} M. Yor, 
{\em Exponential functionals of Brownian motion and related processes} 
(Springer, Berlin, 2000).

\bibitem{2} S. N. Majumdar, 
{\em Brownian functionals in physics and computer science}, 
Current Science {\bf 89}, 2076 (2005).

\bibitem{3} A. Perret, A. Comtet, S. N. Majumdar, and G. Schehr, 
{\em On Certain Functionals of the Maximum of Brownian Motion and Their Applications}, 
J. Stat. Phys. {\bf 161}, 1112 (2015).

\bibitem{4} D. Boyer, D. S. Dean, C. Mej\'{i}a-Monasterio, and G. Oshanin, 
{\em Distribution of the least-squares estimators of a single Brownian trajectory
diffusion coefficient}, 
J. Stat. Mech. \textbf{(2013)} P04017.

\bibitem{4a} D. Boyer, D. S. Dean, C. Mej\'{i}a-Monasterio, and G. Oshanin,
{\em Optimal estimates of the diffusion coefficient of a single Brownian trajectory},  
Phys. Rev. E {\bf 85}, 031136 (2012).  

\bibitem{Borodin} A. N. Borodin and P. Salminen,
{\em Handbook of Brownian Motion: Facts and Formulae},
(Birkh{\"a}user Verlag, Basel, 1996).

\bibitem{comb1} Arkhincheev V E and Baskin E M 1991
Anomalous diffusion and drift in a comb model of percolation clusters 
{\em Sov. Phys. JETP} {\bf 73}, 161.

\bibitem{comb2} Sandev T, Iomin A, Kantz H, Metzler R and Chechkin A 2016
Comb model with slow and Ultraslow diffusion
{\em Math. Model. Nat. Phenom.} {\bf 11}, 18. 

\bibitem{arcsine} P. L\'evy, \emph{Sur certains processus stochastiques
homog{\`e}nes}, Compositio Mathematica \textbf{7}, 283 (1939).

\bibitem{51} H. Geman and M. Yor,
{\em Bessel processes, Asian options, and perpetuities}, 
Math. Finance {\bf 3}, 349 (1993).

\bibitem{61} P. Wilmott, J. Dewynne, and S. Howison, 
{\em Option Pricing: Mathematical Models and Computation},
(Oxford Financial Press, 2000).

\bibitem{71} G. Oshanin and G. Schehr,
{\em Two stock options at the races: Black-Scholes forecasts}, 
Quant. Finance {\bf 12}, 1325 (2012).

\bibitem{peters} O. Peters and W. Klein,
\emph{Ergodicity breaking in geometric Brownian motion},
Phys. Rev. Lett. \textbf{110}, 100603 (2013).

\bibitem{deepak} A. G. Cherstvy, D. Vinod, E. Aghion, A. V. Chechkin, and R. Metzler,
\emph{Time averaging, ageing and delay analysis of financial time series},
New J. Phys. \textbf{19}, 063045 (2017).

\bibitem{100} S. F. Burlatsky, G. Oshanin, A. Mogutov, and M. Moreau, 
{\em Non-Fickian steady flux in a one-dimensional Sinai-type disordered system}, 
Phys. Rev. A {\bf 45}, R6955 (1992).

\bibitem{101} G. Oshanin, A. Mogutov and M. Moreau, 
{\em Steady flux in a continuous-space Sinai chain}, 
J. Stat. Phys. {\bf 73}, 379 (1993).

\bibitem{102} C. Monthus and A. Comtet,
{\em On the flux distribution in a one dimensional disordered system}, 
J. Phys. I. France {\bf 4}, 635 (1994).

\bibitem{103} A. Comtet, C. Monthus, and M. Yor, 
{\em Exponential functionals of Brownian motion and disordered systems}, 
J. Appl. Probab. {\bf 35}, 255 (1998).

\bibitem{104} G. Oshanin, A. Rosso, and G. Schehr, 
{\em Anomalous fluctuations of currents in Sinai-type random chains with strongly
correlated disorder},
Phys. Rev. Lett. {\bf 110}, 100602 (2013).

\bibitem{104a} A. G. Cherstvy, A. V. Chechkin, and R. Metzler,
\emph{Anomalous
diffusion and ergodicity breaking in heterogeneous diffusion processes},
New J. Phys. \textbf{15}, 083039 (2013).

\bibitem{104b} A. G. Cherstvy and R. Metzler,
\emph{Non-ergodicity, fluctuations, and criticality in heterogeneous
diffusion processes},
Phys. Rev. E \textbf{90}, 012134 (2014).

\bibitem{105} G. Oshanin and S. Redner, 
{\em  Helix or coil? Fate of a melting heteropolymer}, 
Europhys. Lett. {\bf 85}, 10008 (2009).

\bibitem{pccp} R. Metzler, J.-H. Jeon, A. G. Cherstvy, and E. Barkai,
\emph{Anomalous diffusion models and their properties: non-stationarity,
non-ergodicity, and ageing at the centenary of single particle tracking},
Phys. Chem. Chem. Phys. \textbf{16}, 24128 (2014).

\bibitem{elipt} F. D. Stefani, J. P. Hoogenboom, and E. Barkai,
\emph{Beyond quantum jumps: Blinking nanoscale light emitters},
Phys. Today \textbf{62(2)}, 34 (2009).

\bibitem{greb3} Y. Lanoisel{\'e}e and D. S. Grebenkov, \emph{Revealing nonergodic
dynamics in living cells from a single particle trajectory},
Phys. Rev. E \textbf{93}, 052146 (2016).

\bibitem{alek} A. Weron, J. Janczura, E. Boryczka, T. Sungkaworn, and D. Calebiro,
\emph{Statistical testing approach for fractional anomalous diffusion classification},
Phys. Rev. E \textbf{99}, 042149 (2019).

\bibitem{ralf} S. Bo, F. Schmidt, R. Eichhorn, and G. Volpe,
\emph{Measurement of anomalous diffusion using recurrent neural networks},
Phys. Rev. E \textbf{100}, 010102 (2019).

\bibitem{samu} S. Thapa, M. A. Lomholt, J. Krog, A. G. Cherstvy, and R. Metzler,
\emph{Bayesian nested sampling analysis of single particle tracking data: maximum
likelihood model selection applied to stochastic diffusivity data},
Phys. Chem. Chem. Phys. \textbf{20}, 29018 (2018).

\bibitem{samu1} A. G. Cherstvy, S. Thapa, C. E. Wagner, and R. Metzler,
\emph{Non-Gaussian, non-ergodic, and non-Fickian diffusion of tracers in mucin
hydrogels},
Soft Matter \textbf{15}, 2526 (2019).

\bibitem{jae_jpa} J.-H. Jeon and R. Metzler, \emph{Analysis of short subdiffusive
time series: scatter of the time averaged mean squared displacement},
J. Phys. A \textbf{43}, 252001 (2010).

\bibitem{greb4} D. S. Grebenkov, \emph{Probability Distribution of the
Time-Averaged Mean-Square Displacement of a Gaussian Process}, Phys. Rev. E
\textbf{84}, 031124 (2011).

\bibitem{greb5} A. Andreanov and D. S. Grebenkov, \emph{Time-averaged MSD of
Brownian motion}, J. Stat. Mech. \textbf{(2012)}, P07001.

\bibitem{jakub} J. \'Sl\c{e}zak, R. Metzler, and M. Magdziarz,
\emph{Codifference can detect ergodicity breaking and non-Gaussianity},
New J. Phys. \textbf{21}, 053008 (2019).

\bibitem{blac73} F. Black and M. Scholes, \emph{The pricing of options
and corporate liabilities}, J. Polit. Econ. \textbf{81}, 637 (1973).

\bibitem{mert71} R. C. Merton, \emph{Optimum consumption and portfolio
rules in a continuous-time model}, J. Econ. Theory \textbf{3}, 373 (1971).

\bibitem{mert75} R. C. Merton, \emph{Option pricing when underlying
stock returns are discontinuous,} J. Financ. Econ. \textbf{3}, 125 (1976).

\bibitem{heston} S. L. Heston, A Closed-Form Solution for Options
with Stochastic Volatility with Applications to Bond and Currency Options,
Rev. Financial Studies \textbf{6}, 327 (1993).

\bibitem{cir} J. C. Cox and S. A. Ross, \emph{The valuation of options
for alternative stochastic processes}, J. Finan. Econom. \textbf{3}, 145
(1976).

\bibitem{cir1} J. C. Cox, J. E. Ingersoll, and S. A. Ross, \emph{A
Theory of the term structure of interest rates}, Econometrica \textbf{53},
385 (1985).

\bibitem{yako} A. A. Dragulescu and V. M. Yakovenko, \emph{Probability
distribution of returns in the Heston model with stochastic volatility},
Quant. Finance \textbf{2}, 443 (2002).

\bibitem{carlos} C. Mej\'{i}a-Monasterio, G. Oshanin, and G. Schehr,
\emph{First passages for a search by a swarm of independent random searchers},
J. Stat. Mech. P06022 (2011).

\bibitem{carlos1} T. Mattos, C. Mej\'{i}a-Monasterio, R. Metzler, and G. Oshanin,
\emph{First passages in bounded domains: When is the mean first passage
time meaningful?},
Phys. Rev. E {\bf 86}, 031143 (2012).

\bibitem{aljaz} A. Godec and R. Metzler,
\emph{Universal proximity effect in target search kinetics in the few encounter
limit},
Phys. Rev. X {\bf 6}, 041037 (2016).

\bibitem{denis} D. Grebenkov, R. Metzler, and G. Oshanin,
\emph{Strong defocusing of molecular reaction times: geometry and reaction control},
Comm. Chem. \textbf{1}, 96 (2018).

\end{thebibliography}

\end{document}